\documentclass{ws-ijmpa}
\usepackage[super,compress]{cite}
\usepackage{graphicx}
\usepackage{braket}
\usepackage{bm}
\usepackage{comment}
\usepackage{here}
\usepackage{multirow}
\usepackage{color}

\def\Journal#1#2#3#4{{#1} {\bf #2}, #3 (#4)}
 	
\def\ARNPS{Annu. Rev. Nucl. Part. Sci.} 
\def\AandA{Astron. Astrophys.} 

\def\APJ{Astrophys. J.}

\def\CMP{Commn. Math. Phys.}

\def\CQG{Class. Quantum Grav.}

\def\EPJC{Eur. Phys. J. C}
\def\EPJP{Eur. Phys. J. Plus}

\def\IJMPA{Int. J. Mod. Phys. A}
\def\IJMPD{Int. J. Mod. Phys. D}

\def\JCAP{J. Cosmol. Astropart. Phys.}
\def\JHEP{J. High Energy Phys.}

\def\MNRAS{Mon. Not. R. Astron. Soc.}
\def\MPLA{Mod. Phys. Lett. A}

\def\PDU{Phys. Dark. Univ.}

\def\PLB{{Phys. Lett.} B}

\def\PPNP{Prog. Part. Nucl. Phys.}

\def\PRL{Phys. Rev. Lett.}
\def\PRD{Phys. Rev. D}

\def\PTP{Prog. Theor. Phys.}
\def\PTEP{Prog. Theor. Exp. Phys.}

\def\RPP{Rep. Prog. Phys.}

\def\UNIV{universe.}

\begin{document}
\markboth{T. Kitabayashi and A. Takeshita}{WIMP/FIMP dark matter and primordial black holes with memory burden effect}

%
\catchline{}{}{}{}{}
%


\title{WIMP/FIMP dark matter and primordial black holes with memory burden effect}

\author{Teruyuki Kitabayashi\footnote{Corresponding author}}

\address{Department of Physics, Tokai University,\\
4-1-1 Kitakaname, Hiratsuka, Kanagawa 259-1292, Japan\\
teruyuki@tokai.ac.jp}

\author{Amane Takeshita}

\address{Graduate School of Science, Tokai University,\\
4-1-1 Kitakaname, Hiratsuka, Kanagawa 259-1292, Japan}

\maketitle

\begin{history}
\received{Day Month Year}
\revised{Day Month Year}
\end{history}

\begin{abstract}
The lifetime of primordial black holes (PBHs), which formed in the early universe, can be extended by the memory burden effect. Light PBHs may exist today and be candidates for dark matter (DM). We assume that DM is made of thermally produced weakly interacting massive particles (WIMPs), WIMPs produced via the Hawking radiation of PBHs, and PBHs that survived Hawking evaporation via the memory burden effect. Feebly interacting massive particles (FIMPs) are alternatives to WIMPs. Focusing on parameter regions where thermal production dominates and PBHs never dominate the energy density of the Universe, we identify a sufficient condition under which DM particles emitted from PBHs do not thermalize with the thermal bath. In this regime, the total DM relic abundance can be consistently obtained as the sum of the three components. In addition, we show that the contribution from gravitational freeze-in via graviton exchange remains subdominant within the parameter space considered.
\end{abstract}



\section{Introduction}
The existence and production mechanism of dark matter (DM) are a long-standing mystery in cosmology and particle physics \cite{Arbey2021PPNP}. The traditional and most studied thermal DM production mechanism is freeze out (FO) \cite{Kolb1990text}. In this mechanism, DM particles, $\chi$, reached thermal equilibrium with thermal bath particles, $f$, in the early universe through annihilation and pair production, $\chi\chi \leftrightarrow \bar{f}f$. Subsequently, the temperature of the bath particles drops below the mass of the DM particles and $\Gamma \lesssim H$ holds, where $\Gamma$ is the reaction rate of annihilation, $\chi\chi \rightarrow \bar{f}f$, and $H$ is the Hubble parameter. Then, the comoving number density of the DM particles remains constant. A typical weak-scale cross section leads to an observed relic abundance of DM, so the FO scenario is usually discussed in the context of weakly interacting massive particles (WIMPs). Although WIMPs are a leading candidate explanation for DM, they cannot be detected in numerous experiments. Thus, feebly interacting massive particles (FIMPs) \cite{Hall2010JHEP,Bernal2017IJMPA} have been proposed based on an idea opposite that of WIMPs. FIMPs are DM particles that never reached thermal equilibrium with thermal bath particles because of very weak coupling between them. Assuming a negligible initial density of FIMPs, the feeble interaction of DM particles with thermal bath particles leads to FIMP production during the expansion of the universe. This DM production scheme, called the freeze-in (FI) mechanism, is applicable to scenarios such as decay ($f_1 \rightarrow f_2 \chi$), scattering ($f_1f_2 \rightarrow f_3\chi$), and pair production ($f_1f_2 \rightarrow \chi\chi$). 

Primordial black holes (PBHs), which formed in the early universe, are also feasible candidates for explaining DM \cite{Carr1975APJ,Carr2020ARNPS,Carr2021RPP,Auffinger2023PPNP,Khlopov2010RAA,Belotsky2014MPLA,Belotsky2019EPJC,Heydari2022EPJC,Heydari2022JCAP,Heydari2024JCAP,Heydari2024EPJC,Heydari2024ApJ,Cheek2022PRD1}. PBHs emit particles through Hawking radiation \cite{Hawking1975CMP}, which reduces their mass. Therefore, PBHs have a finite lifetime. However, if this lifetime is longer than the age of the universe, then some PBHs may not completely evaporate and may exist today.

PBHs that formed with a mass of less than approximately $10^{15}$ g were believed not to have survived to the present day. However, recent findings suggest that the lifetime of PBHs can be extended by the memory burden effect \cite {Dvali2016FortschrPhys,Dvali2020PRD,Alexandre2024PRD,Dvali2024PRD,Thoss2024MNRAS,Chianese2025arXiv}. Quantum corrections for PBHs due to the memory burden effect are being studied extensively to explain the relic abundance of PBHs \cite{Haque2024JCAP,KohriPRD2025,Montefalcone2025arXiv,Dvali2025arXiv}, gravitational waves (GWs) \cite{Bhaumik2024JHEP,Barman2024JCAP09,Barman2024JCAP10,Balaji2024JCAP11,Jiang2024JCAP12,Loc2025PRD,KohriPRD2025,Athron2025JCAP,Barker2025PRD,Gross2025arXiv}, axion DM \cite{Bandyopadhyay2025arXiv}, baryon asymmetry \cite{Borah2024arXiv,Barman2024arXiv,Calabrese2025arXiv}, cosmic neutrinos \cite{Chianese2024arXiv,Chaudhuri2025arXiv,Dondarini2025arXiv,Zantedeschi2025arXiv}, compact stars \cite{Basumatary2025PRD}, and phenomena beyond those in the standard model (SM) of particle physics \cite{Federico2025PRD}. 

We assume that DM consists of WIMPs/FIMPs and survived PBHs. As Hawking radiation is induced by gravity, PBHs evaporate into all particle species. WIMPs/FIMPs are also produced by Hawking radiation. Thus, we consider the following DM components:
\begin{itemize}
\item thermally produced WIMPs/FIMPs
\item WIMPs/FIMPs produced by the Hawking radiation of PBHs
\item survived PBHs
\end{itemize}

Similar studies neglect the memory burden effect \cite{Fujita2014PRD,Gondolo2020PRD, Cheek2022PRD,Chanda2025arXiv,Kitabayashi2021IJMPA,Kitabayashi2022PTEP,Kitabayashi2022PTEP2,Kitabayashi2022IJMPA,Kitabayashi2024PDU,Takeshita2025IJMPA}. Considering the memory burden effect, researchers attempted to explain the relic abundance of DM via the Hawking radiation of PBHs and survived PBHs \cite{Haque2024JCAP}. However, they did not account for DM related to thermally produced particles. 

We simultaneously consider thermally produced WIMPs/FIMPs and the memory burden effect of PBHs to explain the relic abundance of DM. We can omit the contribution of thermally produced WIMPs/FIMPs to the relic abundance of DM if the effect of their thermal production is much smaller than that of particle production by PBHs via Hawking radiation. In this case, the results are the same as those previously reported \cite{Haque2024JCAP}.

According to Ref. \cite{Haque2024JCAP}, we discuss a case where the semiclassical phase of Hawking evaporation ends before big bang nucleosynthesis (BBN) and the PBHs evaporate completely after the present time due to the memory burden effect (the so-called early-matter-dominated era does not exist). To examine a setup different from those in previous studies, we focus on a case where the effect of the thermal production of WIMPs/FIMPs considerably exceeds that of PBH particle production via Hawking radiation.

The remainder of this paper is organized as follows. Section \ref{section:memory_burden_effect} provides a brief review of PBHs with the memory burden effect. In Section \ref{section:constraints}, we discuss the constraints on PBHs from BBN, cosmic microwave background radiation (CMB), GWs, and warm DM (WDM). In Section \ref{section:PBH_WIMP_FIMP}, we assess the combined effects of PBHs and thermally produced WIMPs/FIMPs on the relic abundance of DM. Finally, Section \ref{section:summary} provides a summary of this study and its findings.

\section{PBH with memory burden effect \label{section:memory_burden_effect}}
\subsection{Time evolution}
Neglecting the memory burden effect, the time evolution of PBH mass $M_{\rm BH}$ is given by 
\begin{equation}
\frac{dM_{\rm BH}}{dt} = -\epsilon\frac{M_{\rm P}^4}{M_{\rm BH}^2},
\label{Eq:dMdt_without_memory_burden}
\end{equation}
where $M_{\rm P} \simeq 2.4\times 10^{18}$ GeV is the reduced Planck mass and
\begin{equation}
\epsilon = \frac{27}{4}\frac{\pi g_\ast(T_{\rm BH})}{480},
\end{equation}
with $g_\ast (T_{\rm BH})$ being the effective number of degrees of freedom. The Hawking temperature, $T_{\rm BH}$, is defined as
\begin{equation}
T_{\rm BH}= \frac{M_{\rm P}^2}{M_{\rm BH}}.
\end{equation}
After integrating Eq. (\ref{Eq:dMdt_without_memory_burden}) from the time of PBH formation, $t_{\rm in}$, to time $t$, the PBH mass at time $t$ is calculated as
\begin{equation}
M_{\rm BH} = M_{\rm in} \left[ 1- \Gamma_{\rm BH}^0 (t-t_{\rm in})\right]^{1/3},
\label{Eq:MBH_without_memory_burden}
\end{equation}
where $M_{\rm in}$ is the initial PBH mass. The decay width associated with PBH evaporation is estimated to be
\begin{equation}
\Gamma_{\rm BH}^0 = \frac{3\epsilon M_{\rm P}^4}{M_{\rm in}^3}.
\label{Eq:Gamma0}
\end{equation}
Assuming PBH formation in a radiation-dominated era, PBH lifetime $\tau^0$ is obtained as
\begin{equation}
\tau^0 = t_{\rm ev}^0 - t_{\rm in} = \frac{1}{\Gamma_{\rm BH}^0} \simeq 2.4\times 10^{-28} \left(\frac{M_{\rm in}}{1 {\rm g}} \right)^3 \ {\rm s},
\label{Eq:tau0}
\end{equation}
by solving $M_{\rm BH}(t_{\rm ev}^0)=0$, where $t_{\rm ev}^0$ is the evaporation time. The PBHs with $M_{\rm in} \lesssim 10^9$ g have evaporated before BBN, we use $t_{\rm BBN} = 1$ s, whereas PBHs with $M_{\rm in} \gtrsim 10^{15}$ g can survive until today without memory burden effect.

Considering the memory burden effect, the time evolution of the PBH mass is modified to become \cite{Haque2024JCAP}
\begin{equation}
\frac{dM_{\rm BH}}{dt} = -\frac{\epsilon}{[S(M_{\rm BH})]^k}\frac{M_{\rm P}^4}{M_{\rm BH}^2},
\label{Eq:dMdt}
\end{equation}
where 
\begin{equation}
S=\frac{1}{2}\left( \frac{M_{\rm BH}}{M_{\rm P}}\right)^2 = \frac{1}{2}\left( \frac{M_{\rm P}}{T_{\rm BH}}\right)^2
\end{equation}
is the black hole entropy. The parameter $k$ characterizes the efficiency of the back reaction effect, with $k=0$ indicating Hawking radiation without the memory burden effect. 

The semiclassical era is assumed valid until the PBH mass reaches 
\begin{equation}
M_{\rm BH}= q M_{\rm in},
\label{Eq:MbhQMin}
\end{equation}
where $0 < q < 1$. From Eq. (\ref{Eq:MBH_without_memory_burden}), we define the final time for the semiclassical era as 
\begin{equation}
t_q = t_{\rm ev}^0 (1-q^3).
\label{Eq:tq}
\end{equation}
After integrating Eq. (\ref{Eq:dMdt}) from time $t_q$ to $t$, we obtain
\begin{equation}
M_{\rm BH} = q M_{\rm in} \left[ 1- \Gamma_{\rm BH}^k (t-t_q)\right]^{1/(3+2k)},
\label{Eq:MBH}
\end{equation}
where
\begin{equation}
\Gamma_{\rm BH}^k = 2^k(3+2k)\epsilon M_{\rm P} \left( \frac{M_P}{qM_{\rm in}} \right)^{3+2k}
\label{Eq:Gamma_k}
\end{equation}
at $k \neq 0$ and $0 < q < 1$. We estimate the evaporation time by solving $M_{\rm BH}(t_{\rm ev}^k)=0$ as \cite{Alexandre2024PRD,Haque2024JCAP,Borah2024arXiv} 
\begin{eqnarray}
t_{\rm ev}^k &=& t_q + \frac{1}{\Gamma_{\rm BH}^k} \simeq  (1-q^3) \frac{M_{\rm in}^3}{3\epsilon M_{\rm P}^4} \nonumber \\
&&+ \frac{1}{2^k(3+2k)\epsilon M_{\rm P}} \left( \frac{qM_{\rm in}}{M_P} \right)^{3+2k}.
\label{Eq:tevk}
\end{eqnarray}
Approximation ($t_{\rm ev}^k \simeq  1/\Gamma_{\rm BH}^k$) is adequate in some cases. For example, for $k=1$ and $q=1/2$, the PBH with $M_{\rm in} \gtrsim 2\times 10^7$ g can survive until today due to the memory burden effect.

\subsection{Initial density}
The initial density of a PBH is usually characterized by the dimensionless parameter
\begin{equation}
\beta = \frac{\rho_{\rm BH}^{\rm in}}{\rho_{\rm r}^{\rm in}},
\end{equation}
where $\rho_{\rm BH}^{\rm in}$ and $\rho_{\rm r}^{\rm in} = \pi^2g_\ast(T_{\rm in}) T_{\rm in}^4/30$ are the initial energy densities of the PBH and radiation, respectively. We assume that the PBH formed during a radiation-dominated era in the universe. Hence, the radiation energy density varies as
\begin{equation}
\rho_{\rm r} (a) = \rho_{\rm r}^{\rm in} \left( \frac{a_{\rm in}}{a} \right)^4.
\end{equation}

As a PBH behaves as matter, its energy density evolves as
\begin{equation}
\rho_{\rm BH} (a) = \rho_{\rm BH}^{\rm in} \left( \frac{a_{\rm in}}{a} \right)^3 
\times \begin{cases}
1 & {\rm for} \  a < a_{\rm q}  \\
q  & {\rm for} \  a > a_{\rm q}  \\
\end{cases},
\end{equation}
where $a$ is a scale factor and instant mass reduction is assumed as $M_{\rm in} \rightarrow q M_{\rm in}$ at $a = a_{\rm q}$. 
According to $\rho_{\rm r} = \rho_{\rm BH}$, the radiation--PBH equivalent may occur at $a_{\rm BH} = a_{\rm in}/(q\beta)$. A PBH-dominant era is possible if $a_{\rm BH} < a_{\rm ev}$, which corresponds to 
\begin{equation}
\beta > \frac{a_{\rm in}}{q a_{\rm ev}} = \beta_{\rm c},
\end{equation}
where
\begin{equation}
\beta_{\rm c} = \frac{1}{q^{5/2+k}}\left(\frac{M_{\rm P}}{M_{\rm in}} \right)^{1+k} \sqrt{\frac{(3+2k)2^k\epsilon}{8\pi \gamma}},
\end{equation}
where $\gamma \simeq 0.2$. We assume that the initial PBH mass is related to the horizon size during PBH formation as 
\begin{equation}
M_{\rm in} = \frac{4}{3}\pi \gamma \rho_{\rm r}^{\rm in} H_{\rm in}^{-3}.
\label{Eq:Min}
\end{equation}
In addition, we use relations $H=1/(2t)$, $a \propto t^{1/2}$, and $H_{\rm in}^2 = \rho_{\rm r}^{\rm in}/(3M_{\rm P}^2)$ for a radiation-dominated universe and set $t_{\rm ev}^k \simeq 1/\Gamma_{\rm BK}^k$ using Eq. (\ref{Eq:Gamma_k}).

\subsection{Particle production}
An origin of the DM is an evaporating PBHs. According to Ref. \cite{Haque2024JCAP}, we divide the era of Hawking radiation into two phases as follows:
\begin{description}
\item[ Phase-I:] Semiclassical evaporation phase.
\item[ Phase-II:] The second phase where quantum correction is effective.
\end{description}

Let $N_{1i}$ and $N_{2i}$ be the number of particles which are emitted from PBH in phase-I and phase-II, respectively. The emission rate of particle species $i$ with mass $m_i$ and internal degrees of freedom $g_i$ per unit time per unit energy due to Hawking radiation is given by \cite{Haque2024JCAP}
\begin{align}
\frac{d^2 N_{1i}}{dEdt} & = \frac{27}{4}\frac{g_i}{32\pi^3}\frac{(E/T_{\rm BH})^2} {\exp (E/T_{\rm BH}) \pm 1},  \\
\frac{d^2 N_{2i}}{dEdt} & = \frac{1}{\left[ S(M_{\rm BH}) \right]^k} \frac{d^2 N_{1i}}{dEdt}, 
\end{align}
where the sign $\pm$ is used for fermions and bosons, respectively. After integrating for all energy mode, we obtain
\begin{align}
\label{Eq:dN_1i_dt}
\frac{dN_{1i}}{dt} & = \frac{27}{4}\frac{\xi g_i \zeta (3)}{16\pi^3}\frac{M_{\rm P}^2} {M_{\rm BH}}, \\
\label{Eq:dN_2i_dt}
\frac{dN_{2i}}{dt} & = \frac{27}{4}\frac{\xi g_i \zeta (3) 2^k}{16\pi^3}\frac{M_{\rm P}^{2+2k}} {M_{\rm BH}^{1+2k}}, 
\end{align}
where $\xi = 1$ for bosons and $\xi = 3/4$ for fermions.

The total number of particle species $i$ from evaporation of BH is controlled by the relation between the particle mass $m_i$ and the PBH formation temperature
\begin{equation}
T_{\rm in}=T_{\rm BH}(t_{\rm in}) = \frac{M_{\rm P}^2}{M_{\rm in}}.
\label{Eq:Tin}
\end{equation}
If $m_i < T_{\rm in}$, the particle emission from PBH occurs from $t_{\rm in}$ to $t_{\rm ev}$. On the other hand, if $m_i > T_{\rm in}$, the particle emission will start from a time where $m_i = T_{\rm BH}(t_i)$. From Eqs. (\ref{Eq:MBH_without_memory_burden}), (\ref{Eq:MBH}), and (\ref{Eq:Gamma_k}), we obtain
\begin{equation}
t_i = 
\begin{cases}
t_{\rm ev}^0 \left[1 - \left(\frac{M_{\rm P}^2}{M_{\rm in} m_i}\right)^3 \right]  & {\rm for} \  t_i < t_{\rm q} \  ({\rm phase-I} ) \\
t_{\rm ev}^k \left[1 - \left(\frac{M_{\rm P}^2}{q M_{\rm in} m_i}\right)^{3+2k} \right]  & {\rm for} \  t_i > t_{\rm q} \  ({\rm phase-II} ) \\
\end{cases}
\end{equation}
with assumption of $t_i \gg t_{\rm in}$.

For $m_i < T_{\rm in}$, integrating Eqs. (\ref{Eq:dN_1i_dt}) within $[t_{\rm in}, t_{\rm q}]$ and (\ref{Eq:dN_2i_dt}) within $[t_{\rm q}, t]$, the number of particle species $i$ is given by
\begin{align}
N_{1i} & = \frac{15\xi g_i \zeta(3)}{g_\ast(T_{\rm BH})\pi^4} (1-q^2)\frac{M_{\rm in}^2} {M_{\rm P}^2}, \\
N_{2i} & = \frac{15\xi g_i \zeta(3)}{g_\ast(T_{\rm BH})\pi^4} \frac{q^2M_{\rm in}^2} {M_{\rm P}^2} \left[1-\left(1-\frac{t-t_{\rm q}}{t_{\rm ev}} \right)^{2/(3+2k)}\right].
\end{align}

On the other hand, there are two cases for $m_i > T_{\rm in}$: $t_i < t_{\rm q}$ and $t_i > t_{\rm q}$. For $t_i < t_{\rm q}$, integrating Eqs. (\ref{Eq:dN_1i_dt}) within $[t_i, t_{\rm q}]$ and (\ref{Eq:dN_2i_dt}) within $[t_{\rm q}, t]$, the number of particle species $i$ is given by
\begin{align}
N_{1i} & = \frac{15\xi g_i \zeta(3)}{g_\ast(T_{\rm BH})\pi^4} \left( \frac{M_{\rm P}^2} {m_i^2} - \frac{q^2M_{\rm in}^2} {M_{\rm P}^2}\right), \\
N_{2i} & = \frac{15\xi g_i \zeta(3)}{g_\ast(T_{\rm BH})\pi^4} \frac{q^2M_{\rm in}^2} {M_{\rm P}^2} \left[1-\left(1-\frac{t-t_{\rm q}}{t_{\rm ev}} \right)^{2/(3+2k)}\right].
\end{align}
For $t_i > t_{\rm q}$, corresponding to $m_i > M_{\rm P}^2 / (q M_{\rm in}) = T_{\rm in}/q$, we obtain
\begin{align}
N_{1i} & = 0, \\
N_{2i} & = \frac{15\xi g_i \zeta(3)}{g_\ast(T_{\rm BH})\pi^4} \frac{q^2M_{\rm in}^2} {M_{\rm P}^2} \left[\left(1-\frac{t_i-t_{\rm q}}{t_{\rm ev}} \right)^{2/(3+2k)} - \left(1-\frac{t-t_{\rm q}}{t_{\rm ev}} \right)^{2/(3+2k)}\right].
\end{align}
%

\subsubsection{$\beta > \beta_{\rm c}$}
For $\beta > \beta_{\rm c}$, PBHs dominate the Universe before PBH decays. Since a PBH behaves as matter ($H = 2/(3t)$) and according to the Friedmann equation, we obtain
\begin{equation}
H^2(a_{\rm ev}) = \frac{\rho_r(a_{\rm ev})}{3M_{\rm P}^2} =\frac{1}{3M_{\rm P}^2}\frac{\pi^2 g_\ast T^4(a_{\rm ev})}{30}  = \frac{4 (\Gamma_{\rm BH}^k)^2}{9}, 
\end{equation}
and
\begin{equation}
T(a_{\rm ev}) = M_{\rm P} \left( \frac{4}{3} \frac{30}{\pi^2 g_\ast }\right)^{1/4} \left[(3+2k)2^k\epsilon \left(\frac{M_{\rm P}}{qM_{\rm in}} \right)^{3+2k} \right]^{1/2}.
\label{Eq:Tev_beta>betac}
\end{equation}

In the PBH dominated era, PBH energy density gets converted into radiation density entirely through the evaporation. The reheating temperature would be the temperature of the decay products \cite{Haque2024JCAP}. If we suppose an instant thermalization, we can use the relation of $\rho_{\rm BH}(a_{\rm ev}) = \rho_{r}(a_{\rm ev}) \simeq n_{\rm BH}(a_{\rm ev}) q M_{\rm in}$. We have
\begin{equation}
n_{\rm BH}(a_{\rm ev})=\frac{4}{3}M_{\rm P}^3(3+2k)^2 2^{2k}\epsilon^2 \left(\frac{M_{\rm P}}{q M_{\rm in}}\right)^{7+4k}.
\end{equation}

The relic abundance of particle DM $\chi$ from PBH evaporation is obtained as \cite{Mambrini2021text}
\begin{equation}
\Omega_\chi^{\rm ev} h^2=1.6 \times 10^8 \frac{g_0}{g_{\rm ev}} \frac{n_\chi(a_{\rm ev})}{T^3(a_{\rm ev})} \frac{m_\chi}{\rm GeV}, 
\end{equation}
where $g_0=3.91$ and $g_{\rm ev} \simeq 106.75$ are the effective number of degrees for entropy density at today and at the end of evaporation, respectively. $n_\chi(a_{\rm ev}) = n_{\rm BH}(a_{\rm ev}) N_\chi$ is the number density of particle DM at the end of evaporation. 
The relic abundance of particle DM emitted by Hawking radiation is given by (Eqs.(3.22) and (3.23) in Ref.\cite{Haque2024JCAP})

\begin{equation}
\Omega_\chi^{\rm ev} h^2=1.9 \times 10^7 N_\chi g^{1/2}_\ast(T_{\rm BH}) \left( 2^k (3+2k) \right)^{1/2} \left( \frac{M_{\rm P}}{q M_{\rm in}} \right)^{5/2+k} \frac{m_\chi}{\rm GeV}, 
\end{equation}
where
\begin{equation}
N_\chi = \frac{15\xi g_\chi \zeta(3)}{g_\ast(T_{\rm BH})\pi^4}
\times 
\begin{cases}
\left( \frac{M_{\rm in}}{M_{\rm P}} \right)^2, & {\rm for} \ m_\chi < T_{\rm in} \\
\left( \frac{M_{\rm P}}{m_\chi}\right)^2, & {\rm for} \ m_\chi > T_{\rm in} \\
\end{cases}.
\end{equation}
We assume $g_\ast (T_{\rm BH}) \simeq 106.75$. For $k \rightarrow 0$ and $q\rightarrow 1$, the expression of $\Omega_\chi^{\rm ev} h^2$ reduces to the standard semi-classical expression \cite{Gondolo2020PRD, Fujita2014PRD}.

\subsubsection{$\beta < \beta_{\rm c}$}
For $\beta < \beta_{\rm c}$, we suppose that the standard Hawking evaporation (phase-I) occurs before BBN and the PBHs evaporate completely after the present time due to the memory burden effect (phase-II)\ cite{Haque2024JCAP}. 

The Hubble parameter at the time of evaporation is given by
\begin{equation}
H^2(a_{\rm ev}) = \frac{\rho_r(a_{\rm ev})}{3M_{\rm P}^2} =\frac{1}{3M_{\rm P}^2}\frac{\pi^2 g_\ast T^4(a_{\rm ev})}{30}  = \frac{(\Gamma_{\rm BH}^k)^2}{4}.
\end{equation}
From this relation, we obtain the evaporation temperature as
\begin{equation}
T(a_{\rm ev}) = M_{\rm P} \left( \frac{3}{4} \frac{30}{\pi^2 g_\ast }\right)^{1/4} \left[(3+2k)2^k\epsilon \left(\frac{M_{\rm P}}{qM_{\rm in}} \right)^{3+2k} \right]^{1/2}.
\label{Eq:Tev_beta<betac}
\end{equation}

The relic abundance of particle DM $\chi$ from PBH evaporation is obtained as \cite{Mambrini2021text}
\begin{equation}
\Omega_\chi^{\rm ev} h^2=1.6 \times 10^8 \frac{g_0}{g_{\rm q}} \frac{n_\chi(a_{\rm q})}{T^3(a_{\rm q})} \frac{m_\chi}{\rm GeV}, 
\end{equation}
where $g_{\rm q} \simeq 106.75$ is the effective number of degrees for entropy density at the end of phase-I. $n_\chi(a_{\rm q})$ is the number density of particle DM at the end of phase-I. With the relation of
\begin{equation}
\frac{n_\chi(a_{\rm q})}{T^3(a_{\rm q})} = \frac{N_{1\chi} n_{\rm BH}(a_{\rm in})}{T_{\rm in}^3} =  \frac{N_{1\chi} \rho_{\rm BH}^{\rm in}}{T_{\rm in}^3M_{\rm in}} =  \frac{N_{1\chi} \beta \rho_r^{\rm in}}{T_{\rm in}^3M_{\rm in}}, 
\end{equation}
and Eq.(\ref{Eq:Min}), the relic abundance of particle DM emitted in the semiclassical phase (phase-I) is given by (Eq.(3.33) in Ref.\cite{Haque2024JCAP})
\begin{equation}
\Omega_\chi^{\rm ev} h^2=1.8 \times 10^8 N_\chi \beta \left( \frac{M_{\rm P}}{M_{\rm in}} \right)^{3/2} \frac{m_\chi}{\rm GeV}, 
\end{equation}
where
\begin{equation}
N_\chi = \frac{15\xi g_\chi\zeta (3)}{g_\ast (T_{\rm BH}) \pi^4} 
\times 
\begin{cases}
(1-q^2) \left( \frac{M_{\rm in}}{M_{\rm P}} \right)^2, & {\rm for} \ m_\chi < T_{\rm in} \\
\left( \frac{M_{\rm P}}{m_\chi}\right)^2 - \left( \frac{qM_{\rm in}}{M_{\rm P}} \right)^2, & {\rm for} \ m_\chi > T_{\rm in} \\
\end{cases}.
\end{equation}
For $q\rightarrow 0$, the expression of $N_\chi$ reduces to the standard semi-classical expression for $m_\chi < T_{\rm in}$ and  $m_\chi > T_{\rm in}$ \cite{Gondolo2020PRD, Fujita2014PRD}.

\subsection{Survived PBH density}
According to Ref.\cite{Haque2024JCAP}, the $\beta < \beta_c$ case is interesting because we assume that PBHs evaporate completely after the present time due to memory burden effect and the survived PBHs until today are also a candidate of DM (there is no ``early" matter dominated era). Hereafter, we use the relations for $\beta < \beta_c$. 

If PBHs evaporate completely after the present time due to the memory burden effect (phase-II). The survived PBHs until today are also a candidate of DM. The density of survived PBHs is obtained as \cite{Mambrini2021text}
\begin{equation}
\Omega_{\rm PBH} h^2=1.6 \times 10^8 \frac{g_0}{g_{\rm q}} \frac{\rho_{\rm BH}(a_{\rm q})}{T^3(a_{\rm q})} \  {\rm GeV^{-1}}.
\end{equation}
With the relation of 
\begin{equation}
\frac{\rho_{\rm BH}(a_{\rm q})}{T^3(a_{\rm q})}  =  \frac{q \rho_{\rm BH}^{\rm in}}{T_{\rm in}^3} =  \frac{q \beta \rho_r^{\rm in}}{T_{\rm in}^3}, 
\end{equation}
and Eq.(\ref{Eq:Min}), we obtain (Eq. (3.37) in Ref.\cite{Haque2024JCAP})
\begin{equation}
\Omega_{\rm PBH} h^2=4.2 \times 10^{26} q \beta \left( \frac{M_{\rm P}}{M_{\rm in}} \right)^{1/2}.
\end{equation}
%

\section{Constraints \label{section:constraints}}
\subsection{BBN}
We focus on a case where the semiclassical phase of PBH evaporation ends before BBN. The initial mass of the PBH, which is partially evaporated until the final time for the semiclassical era, $t_q$, can be obtained from Eqs. (\ref{Eq:Gamma0}), (\ref{Eq:tau0}), and (\ref{Eq:tq}) as 
\begin{equation}
M_{\rm in,t_q} =1.61\times 10^9 \left(\frac{1}{1-q^3}\right)^{1/3}\left(\frac{t_q}{1 {\rm s}} \right)^{1/3} ~ {\rm g},
\end{equation}
where we assume $t_{\rm in} \ll t_{\rm ev}^0$. 

In this study, we set the time of BBN as $t_{\rm BBN} = 1$ s for simplicity. We obtain 
\begin{equation}
M_{\rm in,t_{\rm BBN}} =1.61\times 10^9 \left(\frac{1}{1-q^3}\right)^{1/3} ~ {\rm g}
\end{equation}
at $t_q = t_{\rm BBN}$. If $M_{\rm in} >  M_{\rm in,t_{\rm BBN}}$, then the semiclassical evaporation of PBHs happens after BBN. Because $1 \le (1-q^3)^{-1/3} \le 1.055$ at $0 < q\le 0.9$, $M_{\rm in,t_{\rm BBN}}$ is almost independent of $q$ in this study. We use $M_{\rm in} \le M_{\rm in,t_{\rm BBN}} =1\times 10^9$ g as the BBN constraint, which is the same as that without the memory burden effect \cite{Carr2021RPP,Kohri1999PRD,Kawasaki2000PRD,Carr2010PRD,Fujita2014PRD}. 

\subsection{CMB}
Constraints on inflation impose a lower limit on the PBH mass. Given that the Hubble parameter at a time is less than or equal to the Hubble parameter during inflation, we obtain the lower limit of the initial mass of the PBH as $M_{\rm in} \gtrsim 0.1$ g \cite{Fujita2014PRD,Carr2021RPP}. Because this constraint comes from PBH generation, it is independent of the memory burden effect. Although this constraint depends on the details of gravitational collapse and the inflation scenario, we regard $M_{\rm in} \gtrsim 0.1$ g as the CMB constraint by assuming a standard slow-roll scenario for simplicity \cite{Carr2021RPP}.

\subsection{GWs}
PBHs are potential sources of GWs \cite{Tomita1967PTP,Matarrese1993PRD,Matarrese1994PRL,Matarrese1998PRD,Noh2004PRD,Carbone2005PRD,Nakamura2007PTP,Ananda2007PRD,Baumann2007PRD,Saito2009PRL,Osano2007JCAP,Espinosa2018JCAP,Kohr2018PRD,Domenech2020IJMPD,Domenech2020JCAP,Wang2023arXiv,Choudhury2014PLB,Choudhury2023arXiv,Domenech2021Universe}. PBH formation is accompanied by GWs, and primordial curvature perturbations can induce GWs in the final stage of inflation, immediately after horizon re-entry. 

GWs can be induced by sources other than primordial fluctuations \cite{Inomata2019JCAP,Inomata2020PRD,Papanikolaou2021JCAP}. PBHs are randomly distributed in space according to approximate Poisson statistics immediately after formation. Although PBH gas behaves similarly to pressureless dust, its spatially inhomogeneous distribution causes isocurvature density fluctuations. At a sufficiently large initial fraction of PBHs in the early-matter-dominated (PBH-dominated) era ($\beta > \beta_{\rm c}$ case), the initial isocurvature fluctuations become curvature perturbations. These curvature perturbations then act as sources of secondary GWs \cite{Inomata2019JCAP,Inomata2020PRD,Inomata2019PRD,Papanikolaou2021JCAP,Domenech2021JCAP,Domenech2021PLB,Bhaumik2022JHEP,Borah2023JHEP,KawasakiArXiv2023}. These induced GWs can be detected via future GW observations, such as those obtained using Hanford--Livingstone--Virgo--KAGRA \cite{LIGO2010CQG,LIGO2015CQG,VIRGO2015CQG,KAGRA2012CQG}, the Einstein Telescope \cite{Punturo2010CQG,Maggiore2020JCAP}, Cosmic Explorer \cite{LIGO2017CQG}, Laser Interferometer Space Antenna \cite{LISA2017arXiv}, Deci-Hertz Interferometer Gravitational-Wave Observatory \cite{Seto2001PRL,Kawamura2006CQG}, and Big Bang Observer \cite{Crowder2005PRD,Harry2006CQG}.

GWs induced by PBHs with the memory burden effect have been extensively studied recently \cite{Bhaumik2024JHEP,Barman2024JCAP09,Barman2024JCAP10,Balaji2024JCAP11,Jiang2024JCAP12,Loc2025PRD,KohriPRD2025,Athron2025JCAP,Barker2025PRD,Gross2025arXiv}. Findings suggest that the GW spectrum amplitude is enhanced by the longer early PBH-dominated phase caused by the memory burden effect, assuming the existence of the early-matter-dominated era ($\beta > \beta_{\rm c}$). 

In this study, PBHs are assumed to evaporate completely after the present time due to the memory burden effect and PBHs that have survived until the present day are regarded as DM candidates ($\beta < \beta_{\rm c}$ should hold). We assume that the early-matter-dominated era did not exist and PBHs never dominated the energy density in the history of the universe. In addition, we focus on a case where the effect of the thermal production of WIMPs/FIMPs substantially exceeds that of particle production by PBHs. In this case, GWs from PBHs almost behave as if the memory burden effect is nonexistent, so a small GW amplitude is expected (see $q=0$ curve in Fig. 3 in Ref. \cite{Bhaumik2024JHEP} for example).

\subsection{WDM}
DM should be cold enough to avoid erasing small-scale structures via free streaming. A constraint on WDM or noncold DM is obtained via the study of the Lyman-$\alpha$ forest \cite{Hui1997ApJ,Gnedin2002MNRAS,Baldes2020JCAP,Boyarsky2009JCAP,Fujita2014PRD,Baur2017JCAP}. 

According to Refs. \cite{Masina2020EPJP,Fujii2002PRD,Barman2024JCAP09}, DM particles produced via PBH evaporation must satisfy a lower bound on their mass,
\begin{eqnarray}
m_{\rm DM} \gtrsim 10^4 \langle E_{\rm DM}(t_{\rm eq})\rangle,
\end{eqnarray}
where $\langle E_{\rm DM}(t_{\rm eq}) \rangle$ denotes the average kinetic energy of DM particles at the epoch of radiation-matter equality. The average kinetic energy at this epoch can be estimated by redshifting the energy at the evaporation time,
\begin{eqnarray}
\langle E_{\rm DM}(t_{\rm eq})\rangle \simeq \langle E_{\rm DM}(t_{\rm ev})\rangle \frac{a_{\rm ev}}{a_{\rm eq}} \simeq \delta T_{\rm BH} \frac{T(a_{\rm eq})}{T(a_{\rm ev})} \left[ \frac{g_{\ast s} (T(a_{\rm eq})) }{g_{\ast s} (T(a_{\rm ev})) } \right]^{1/3}
\end{eqnarray}
Here $T_{\rm BH} \simeq M^2_{\rm P}/(q M_{\rm in})$, and $T(a_{\rm eq}) \simeq 0.75~{\rm eV}$ is the radiation temperature at radiation-matter equality. The function $g_{\ast s}(T)$ represents the effective number of relativistic degrees of freedom contributing to the entropy density.

Following Refs.\cite{Lennon2018JCAP, Masina2020EPJP}, we introduce a numerical factor $\delta \simeq 1.3$ to parametrize the average DM energy at evaporation as $\langle E_{\rm DM}(t_{\rm ev}) \rangle \simeq \delta T_{\rm BH}$. We also assume entropy conservation from the PBH evaporation epoch until radiation-matter equality.

According Eqs. (\ref{Eq:Tev_beta>betac}) and (\ref{Eq:Tev_beta<betac}), the lower bound on the DM mass is given by 
\begin{eqnarray}
m_{\rm DM} &\gtrsim&  g_\ast^{1/4}\left(\frac{qM_{\rm in}}{M_{\rm P}} \right)^{\frac{1+2k}{2}}\frac{1}{\sqrt{ (3+2k)2^k\epsilon}} \left[ \frac{g_{\ast s} (T(a_{\rm eq})) }{g_{\ast s} (T(a_{\rm ev})) } \right]^{1/3} \nonumber \\
&&\times 
\begin{cases}
6.9\times 10^{-6} ~ {\rm GeV} & {\rm for} \ \beta > \beta_{\rm c}\\
7.9\times 10^{-6} ~ {\rm GeV} & {\rm for} \ \beta < \beta_{\rm c}\\
\end{cases}
\nonumber \\
&\gtrsim&  \left(\frac{qM_{\rm in}}{M_{\rm P}} \right)^{\frac{1+2k}{2}}\frac{1}{\sqrt{ (3+2k)2^k}}  
\times
\begin{cases}
3.39\times 10^{-6} ~ {\rm GeV} & {\rm for} \ \beta > \beta_{\rm c}\\
3.88\times 10^{-6} ~ {\rm GeV} & {\rm for} \ \beta < \beta_{\rm c}\\
\end{cases},
\end{eqnarray}
where in the last line, we set $g_{\ast s} (T(a_{\rm eq})) = 3.91$ and $g_\ast = g_{\ast s} (T(a_{\rm ev})) \simeq 106.75$. For example, we obtain
\begin{equation}
m_{\rm DM} \gtrsim \left(\frac{q}{0.5} \right)\left(\frac{M_{\rm in}}{10 ~ {\rm g}}\right)  
\times
\begin{cases}
1.6 ~ {\rm GeV} & {\rm for} \ \beta > \beta_{\rm c}\\
1.9 ~ {\rm GeV} & {\rm for} \ \beta < \beta_{\rm c}\\
\end{cases},
\end{equation}
for $k=1/2$. For a given PBH mass, this bound becomes stronger for larger values of $k$ and $q$ , since in that case PBH evaporation is delayed; as a result, the average kinetic energy of the radiated particles at radiation-matter equality increases \cite{Barman2024JCAP09}.

\section{WIMP/FIMP DM and PBHs \label{section:PBH_WIMP_FIMP}}
\subsection{Thermalization of PBH-produced dark matter}
In general, DM particles injected non-thermally from PBHs may thermalize with the Standard Model plasma if their interaction rate exceeds the Hubble expansion rate, as discussed for example in Ref.\cite{Cheek2022PRD}.

Before proceeding, we clarify under which conditions the DM particles produced via PBH evaporation do not affect the standard FO (or FI) mechanism. In this work, we focus on a restricted parameter region not only characterized by $\beta < \beta_c$, such that PBHs never dominate the energy density of the Universe, but also by
\begin{equation}
\Omega_{\rm FO(FI)} \gg \Omega_{\rm ev},
\end{equation}
so that the contribution of PBH evaporation to the DM abundance is always subdominant. 

We now derive a condition under which DM particles produced via PBH evaporation do not thermalize with the Standard Model plasma. Since the FO condition is defined by $\Gamma_{\rm FO} \simeq H$, it follows immediately that if
\begin{equation}
\Gamma_{\rm ev} \ll H
\end{equation}
is satisfied then the DM particles produced via PBH evaporation do not enter thermal equilibrium, and the relic abundance obtained from the FO mechanism remains unaltered. The interaction rate can be estimated as
\begin{equation}
\Gamma_{\rm ev} \sim n_\chi^{\rm ev}\langle \sigma v \rangle ,
\end{equation}
where $n_\chi^{\rm ev}$ denotes the number density of DM particles injected by PBH evaporation and $\langle \sigma v \rangle$ is the thermally averaged interaction cross section for DM interaction with the Standard Model.

In the case where PBHs never dominate the energy density of the Universe ($\beta < \beta_c$), PBH evaporation takes place during a radiation-dominated era. The number density of DM particles produced by PBHs can then be estimated as
\begin{equation}
n_\chi^{\rm ev}(T) \sim \beta\,\frac{\rho_r(T)}{M_{\rm in}}\,N_\chi ,
\end{equation}
where $\rho_r(T) \sim T^4$. Using the Hubble rate in the radiation-dominated era, $H(T) \sim T^2/M_{\rm P}$, the condition $\Gamma_{\rm ev} \ll H$ can be written as
\begin{equation}
\beta N_\chi \langle \sigma v \rangle  \frac{M_{\rm P}}{M_{\rm in}} \frac{m_\chi^2}{x_{\rm FO}^2} \ll 1,
\label{Eq:Indipendent_condition}
\end{equation}
where we think the FO of WIMPs occurs at \cite{Kolb1990text}
\begin{equation}
x_{\rm FO} =  \frac{m_\chi}{T_{\rm FO}} \simeq 20,
\end{equation}
where $T_{\rm FO}$ is the FO temperature. When the condition in Eq.(\ref{Eq:Indipendent_condition}) is satisfied, the interaction rate of PBH-produced DM particles always remains well below the Hubble expansion rate, and thermalization does not occur. In this case, the relic abundance generated by the FO (or FI) mechanism is not modified by PBH evaporation in the parameter space considered in this work.

\subsection{WIMPs and PBHs}
First, we consider DM made of WIMPs and PBHs. We assume that FO occurs during a radiation-dominated era. The Hubble parameter is related to the radiation temperature and time as $H=(\pi^2 g_\ast / 30)T^2/(3M_{\rm P})=1/(2t)$. Thus, the FO time is 
\begin{equation}
t_{\rm FO} =  \frac{1}{2}\sqrt{\frac{90}{\pi^2 g_\ast}} \frac{M_{\rm P}}{T_{\rm FO}^2}.
\label{Eq:tFO}
\end{equation}

The relic abundance of WIMPs due to FO, $\chi$, is \cite{Gondolo2020PRD} 
\begin{equation}
\Omega_{\rm FO}h^2  = 0.76 \frac{x_{\rm FO}}{g_\ast^{1/2}}\frac{s_0}{M_{\rm P}\rho_{\rm c} \langle \sigma v \rangle} h^2,
\end{equation}
where $\langle \sigma v \rangle$ is the thermally averaged cross section at $\chi\chi \rightarrow \bar{f}f$ and $s_0 = 2.89\times 10^3 \ {\rm cm^{-3}}$ is the entropy density. Considering the so-called WIMP miracle, we parameterize
\begin{equation}
 \langle \sigma v \rangle = \frac{\alpha_{\rm WIMP}^2}{m_\chi^2},
\end{equation}
where $\alpha_{\rm WIMP}^2$ is the effective coupling at $\chi\chi \leftrightarrow \bar{f}f$. The relic abundance of WIMPs is 
\begin{align}
\Omega_{\rm FO}h^2 = 1.7 \times 10^{-10} \frac{1}{\alpha_{\rm WIMP}^2} \left( \frac{ m_\chi}{\rm GeV} \right)^2.
\label{Eq:OmegaFOh2}
\end{align}
If the FO mechanism alone produces DM, then the observed relic abundance, $\Omega_{\rm DM}h^2=0.12$ \cite{Planck2018AA}, is consistent with $\alpha_{\rm WIMP}\simeq 0.026$ and $m_\chi \simeq 1$ TeV.

We assume that the DM particle is a Majorana fermion (e.g., $\xi = 3/4$ and $g_\chi = 1$). In addition, we omit the negligible contribution of $g_\chi = 1$ to the effective number of degrees of freedom and use $g_*=106.75$. 

\begin{figure}[t]
\centering
\includegraphics[keepaspectratio, scale=0.35]{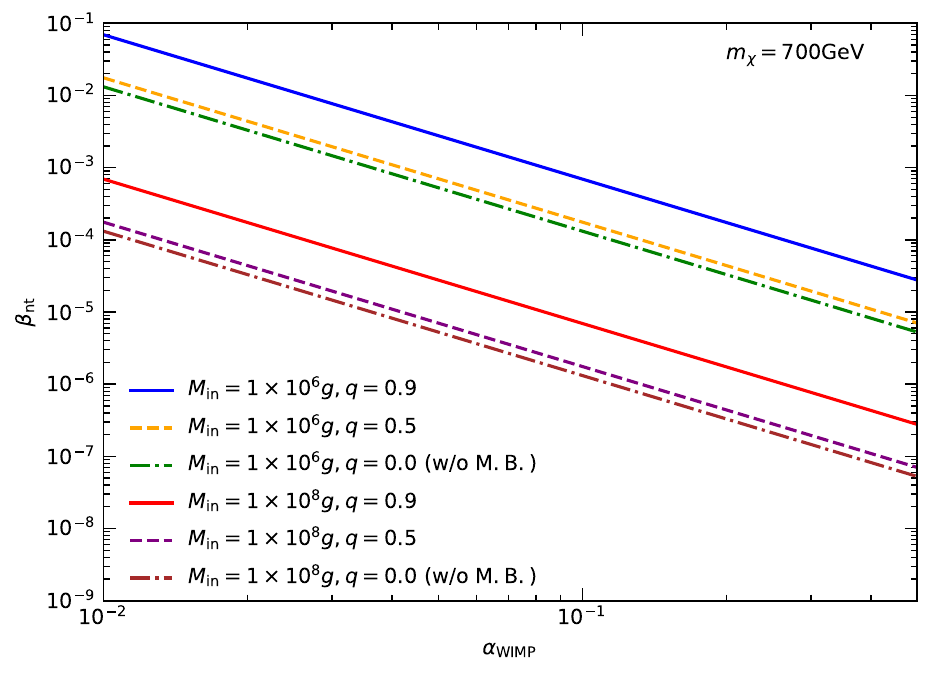}
 \caption{Dependence of the threshold $\beta_{\rm nt}$, which characterizes the condition under which particle DM produced from PBHs does not thermalize, on the coupling strength $\alpha_{\rm WIMP}$. The upper three curves correspond to the case with an initial PBH mass of $10^{6}\,\mathrm{g}$, while the lower three curves correspond to $10^{8}\,\mathrm{g}$. For a fixed PBH mass, the differences among the three curves reflect different choices of the parameter $q$.}
 \label{Fig:WIMP-non-thermal}
 \end{figure}

First, we assume $k=1$ as the simplest case in the following numerical calculations. We regard $\alpha_{\rm WIMP} = 0.0323$ and $m_\chi = 700$ GeV as the benchmark. In this case, we obtain $\Omega_{\rm FO}h^2=0.08$. 

From Eq.(\ref{Eq:Indipendent_condition}), we obtain the non-themalization condition 
\begin{equation}
\beta N_\chi  \alpha_{\rm WIMP}^2  \frac{M_{\rm P}}{M_{\rm in}} \frac{1}{x_{\rm FO}^2} \ll 1,
\label{Eq:Indipendent_condition_WIMP}
\end{equation}
for WIMPs. This inequality shows that non-thermalization of PBH-produced WIMPs can be guarantee for sufficiently small initial PBH density, small interactions, or large initial PBH mass. Also the non-thermalization threshold depends on WIMP mass, initial mass of PBH and/or, strength of the memory burden effect ($N_\chi$ depends on $m_\chi$, $M_{\rm in}$ and/or, $q$). In this study, we focus the case of $\Omega_{\rm FO}h^2 \gg \Omega_{\rm ev}h^2$ (the condition of Eq.(\ref{Eq:Indipendent_condition}) as well as Eq.(\ref{Eq:Indipendent_condition_WIMP}) are satisfied), and PBH-produced WIMPs do not enter thermal equilibrium. The scenario in which PBH-produced WIMPs enter thermal equilibrium lies beyond the scope of the present study. For our purposes, we require not only $\beta < \beta_c$ but also
\begin{equation}
\beta \ll \beta_{\rm nt} = \frac{x_{\rm FO}^2}{N_\chi  \alpha_{\rm WIMP}^2 }\frac{M_{\rm in}}{M_{\rm P}},
\end{equation}
to determine the initial density of PBHs, where $\beta_{\rm nt}$ denotes the threshold for non-thermalization of PBH-origin WIMPs.

Figure~ \ref{Fig:WIMP-non-thermal} illustrates the dependence of the threshold $\beta_{\rm nt}$, which characterizes the condition under which particle DM produced from PBHs does not thermalize, on the coupling strength $\alpha_{\rm WIMP}$. In this figure, the DM mass is fixed to $700~\mathrm{GeV}$. We have checked that varying the DM mass by $\mathcal{O}(50\%)$ does not significantly modify the value of $\beta_{\rm nt}$. The upper three curves correspond to the case with an initial PBH mass of $10^{6}\,\mathrm{g}$, while the lower three curves correspond to $10^{8}\,\mathrm{g}$. For a fixed PBH mass, the differences among the three curves reflect different choices of the parameter $q$. 

We find that, as the memory burden effect becomes stronger (i.e., as $q$ increases), the threshold $\beta_{\rm nt}$ for PBH-produced WIMPs to remain non-thermalized also increases. This is because a larger memory burden effect leads to a slower evaporation of PBHs, which suppresses the total amount of WIMPs emitted even if a larger number of PBHs were formed in the early Universe.

From Fig.\ref{Fig:WIMP-non-thermal}, if we adopt the range $\alpha_{\rm WIMP} = 0.01 \sim 0.5$ as a representative interval for phenomenologically viable WIMP couplings, then for an initial PBH abundance $\beta \ll 10^{-9}$, the condition $\beta \ll \beta_{\rm nt}$ is expected to hold over the PBH mass range and the range of the memory-burden effect considered in this work. As discussed below, the value of $\beta$ required to reproduce the observed DM abundance $\Omega_{\rm DM} h^2 \simeq 0.12$ under our setup, $\Omega_{\rm FO} h^2 \gg \Omega_{\rm ev} h^2$, lies well within this regime. Accordingly, the present analysis focuses on the parameter region in which WIMP originating from PBHs is not expected to thermalize.

Before progress this subsection, we clarify why a numerical solution of the coupled Boltzmann equations is not required in the present analysis. In general, such a treatment becomes necessary only if DM particles injected by PBH evaporation can enter thermal equilibrium with the Standard Model plasma, which requires the interaction rate to satisfy $\Gamma_{\rm ev} \gtrsim H$. Instead of exploring this boundary region, we deliberately restrict ourselves to a parameter space that satisfies a sufficient condition for non-thermalization, $\Gamma_{\rm ev} \ll H$, throughout the cosmological history relevant for FO (or FI).

Using Eq.~(\ref{Eq:Indipendent_condition_WIMP}), this condition can be expressed as $\beta \ll \beta_{\rm nt}$, where $\beta_{\rm nt}$ denotes the threshold initial PBH fraction above which thermalization may occur.  In all benchmark scenarios considered in this work, the hierarchy $\beta / \beta_{\rm nt} \lesssim 10^{-10}$ is realized, placing the system far away from the thermalization boundary.  As a consequence, DM particles emitted by PBHs never reach thermal equilibrium with the thermal bath, and the relic abundance generated by the standard FO (or FI) mechanism remains unchanged.

Solving the coupled Boltzmann equations would therefore not modify the final relic abundance within the parameter space considered here.  A numerical Boltzmann analysis becomes relevant only in scenarios where PBH evaporation injects a sizable DM population close to thermal equilibrium, as studied for instance in Ref.~\cite{Cheek2022PRD}, which are explicitly excluded in the present work.

The present-day density of DM is
\begin{align}
\Omega_{\rm DM}h^2 &=  
\begin{cases}
\Omega_{\rm FO}h^2 + \Omega_{\rm PBH}h^2  & {\rm for} \ t_{\rm ev} ^k< t_{\rm FO} \\
\Omega_{\rm FO}h^2 + \Omega_{\rm PBH}h^2 + \Omega_{\rm ev}h^2  & {\rm for} \ t_{\rm FO} < t_{\rm ev}^k \\
\end{cases}, \nonumber \\
 &= 1.7 \times 10^{-10} \frac{1}{\alpha_{\rm WIMP}^2} \left( \frac{ m_\chi}{\rm GeV} \right)^2  + 4.2 \times 10^{26} q \beta \left( \frac{M_{\rm P}}{M_{\rm in}} \right)^{1/2}   \nonumber \\
& \quad + 
\begin{cases}
0 & {\rm for} \ t_{\rm ev}^k < t_{\rm FO} \\
2.3 \times 10^5 C \beta \frac{m_\chi}{\rm GeV} & {\rm for} \ t_{\rm FO} < t_{\rm ev}^k \\
\end{cases},
\label{Eq:OmegaDMh2FO_and_PBH}
\end{align}
where the first--third terms are due to FO, survived PBHs, and PBH evaporation. The coefficient $C$ is 
\begin{equation}
C = 
\begin{cases}
(1-q^2) \left( \frac{M_{\rm in}}{M_{\rm P}} \right)^{1/2}, & {\rm for} \ m_\chi < T_{\rm in}, \\
\left(\frac{M_{\rm P}}{M_{\rm in}} \right)^{3/2} \left[  \left( \frac{M_{\rm P}}{m_\chi}\right)^2 - \left( \frac{qM_{\rm in}}{M_{\rm P}} \right)^2 \right], & {\rm for} \ m_\chi > T_{\rm in}, \\
\end{cases}
\label{Eq:C}
\end{equation}
where $T_{\rm in}$ is given by Eq. (\ref{Eq:Tin}).

\begin{figure}[t]
\centering
\includegraphics[keepaspectratio, scale=0.35]{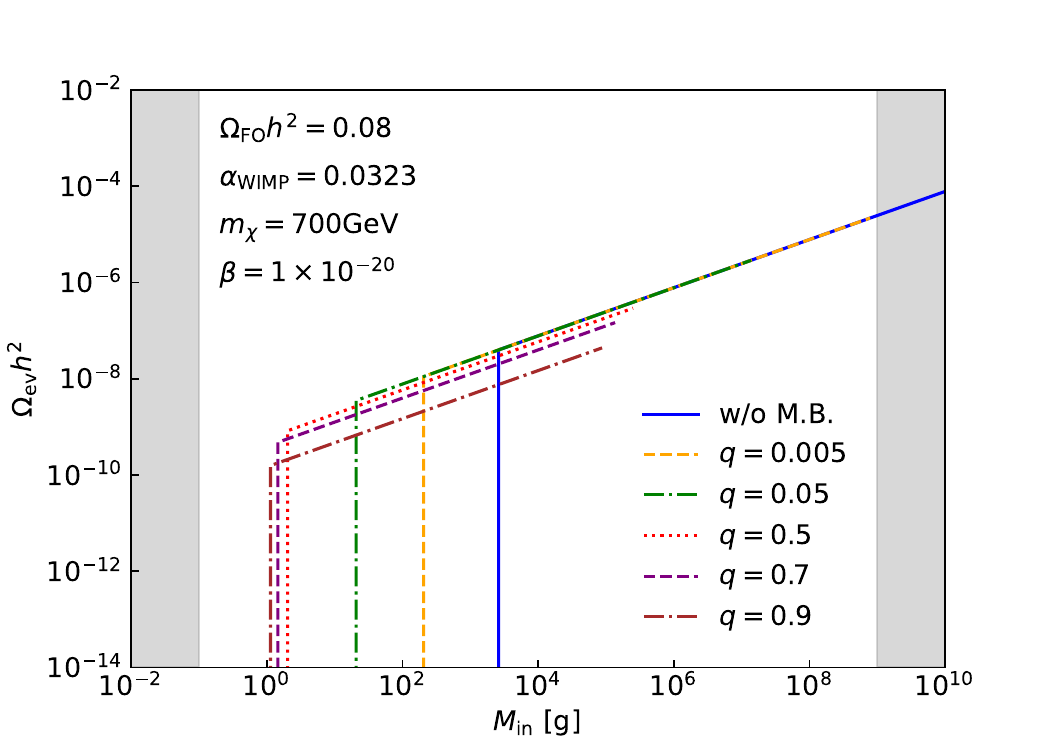}
\includegraphics[keepaspectratio, scale=0.35]{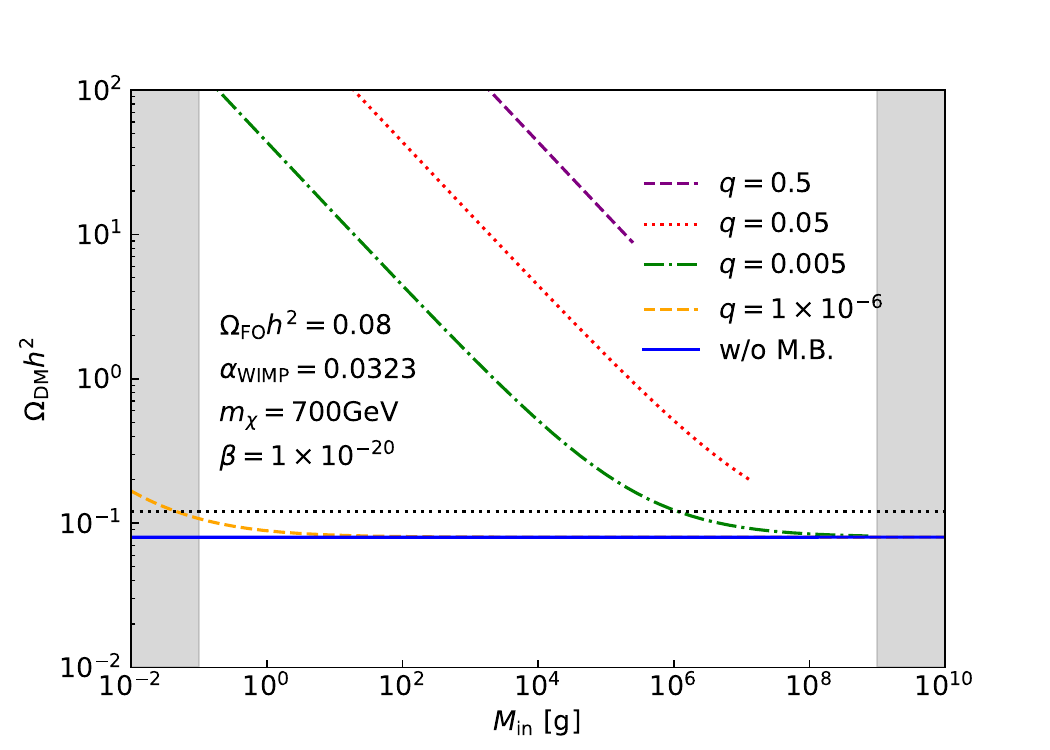}
 \caption{Relic abundance of DM considering WIMPs and PBHs. Left: relic abundance of WIMPs due to Hawking radiation from PBHs, $\Omega_{\rm ev}h^2$. Right: total relic abundance of DM, $\Omega_{\rm DM}h^2$, according to initial PBH mass, $M_{\rm in}$. The left (right) gray region is excluded from the CMB (BBN) constraint.}
 \label{Fig:WIMP-PBH1}
 \end{figure}

The left panel in Fig. \ref{Fig:WIMP-PBH1} shows the relic abundance of WIMPs due to Hawking radiation from PBHs, $\Omega_{\rm ev}h^2$, according to the initial PBH mass, $M_{\rm in}$, at the initial PBH density, $\beta = 1\times 10^{-20}$. The left (right) gray region is excluded from the CMB (BBN) constraint. The blue solid line indicates the prediction without the memory burden effect, and the dotted lines show the predictions with the memory burden effect at $q=0.005-0.9$. $\Omega_{\rm FO}h^2 \gg \Omega_{\rm ev}h^2$ ($\Omega_{\rm ev}/\Omega_{\rm FO} \lesssim 10^{-4}$) holds. As the relic density of WIMPs due to Hawking radiation should vanish at $t_{\rm ev}^k < t_{\rm FO}$, each curve has a lower limit of $M_{\rm in}$. These lower limits can indicate shifts to a lighter mass owing to the memory burden effect via $t_{\rm ev}^k \simeq 1/\Gamma_{\rm BH}^k \propto (q M_{\rm in})^{3+2k}$ from Eqs. (\ref{Eq:Gamma_k}) and (\ref{Eq:tevk}).

The right panel in Fig. \ref{Fig:WIMP-PBH1} shows the total relic abundance of DM, $\Omega_{\rm DM}h^2$, according to the initial PBH mass, $M_{\rm in}$. The blue solid line indicates the prediction without the memory burden effect, and the black horizontal dotted line indicates the observed relic abundance, $\Omega_{\rm DM}h^2=0.12$. The other dotted lines show the predictions with the memory burden effect at various $q$. The relic density of WIMPs due to FO is fixed at $\Omega_{\rm FO}h^2=0.08$, and $\Omega_{\rm DM}h^2 \simeq \Omega_{\rm FO}h^2 + \Omega_{\rm PBH}h^2$ holds. Hence, the trends of the dotted-line curves are reflected in the relic density of the survived PBHs with the memory burden effect, $\Omega_{\rm PBH}h^2 \propto q\beta/M_{\rm in}^{1/2}$. The DM relic abundance is highly sensitive to the memory burden effect. For example, a small memory burden effect of $q \lesssim \mathcal{O}(10^{-3})$ is only allowed at $M_{\rm in} \lesssim \mathcal{O}(10^6)$ g, $\beta=1 \times 10^{-20}$, and $\Omega_{\rm FO}h^2 = 0.08$.

\begin{figure}[t]
\begin{center}
\includegraphics[keepaspectratio, scale=0.35]{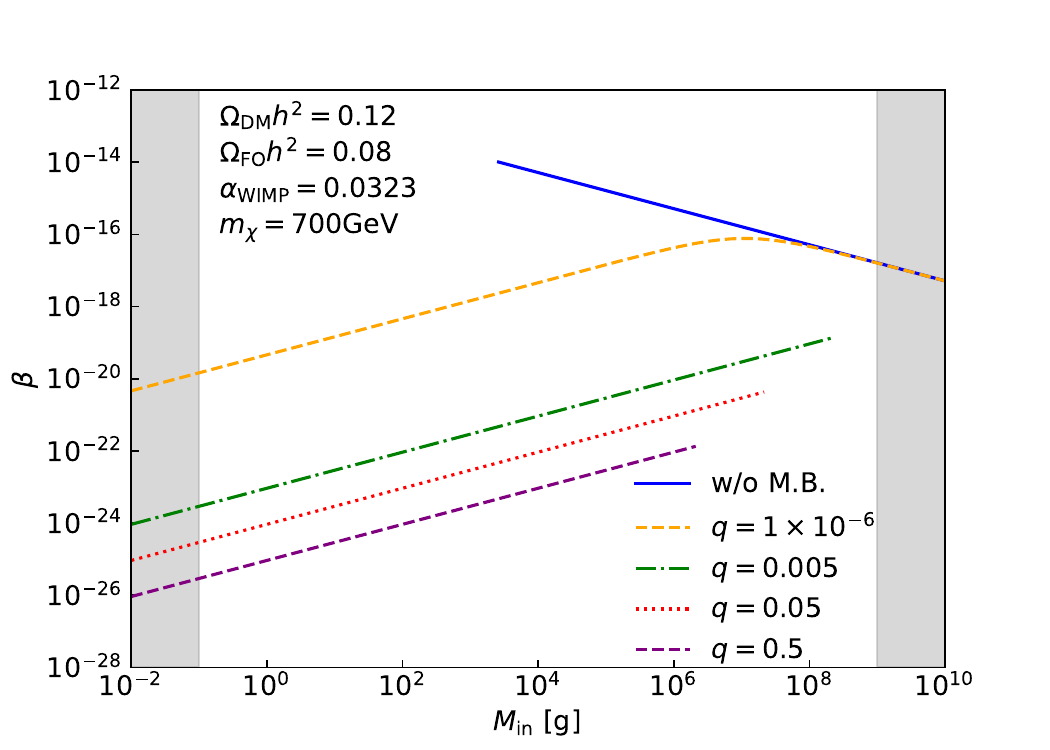}
\end{center}
\includegraphics[keepaspectratio, scale=0.35]{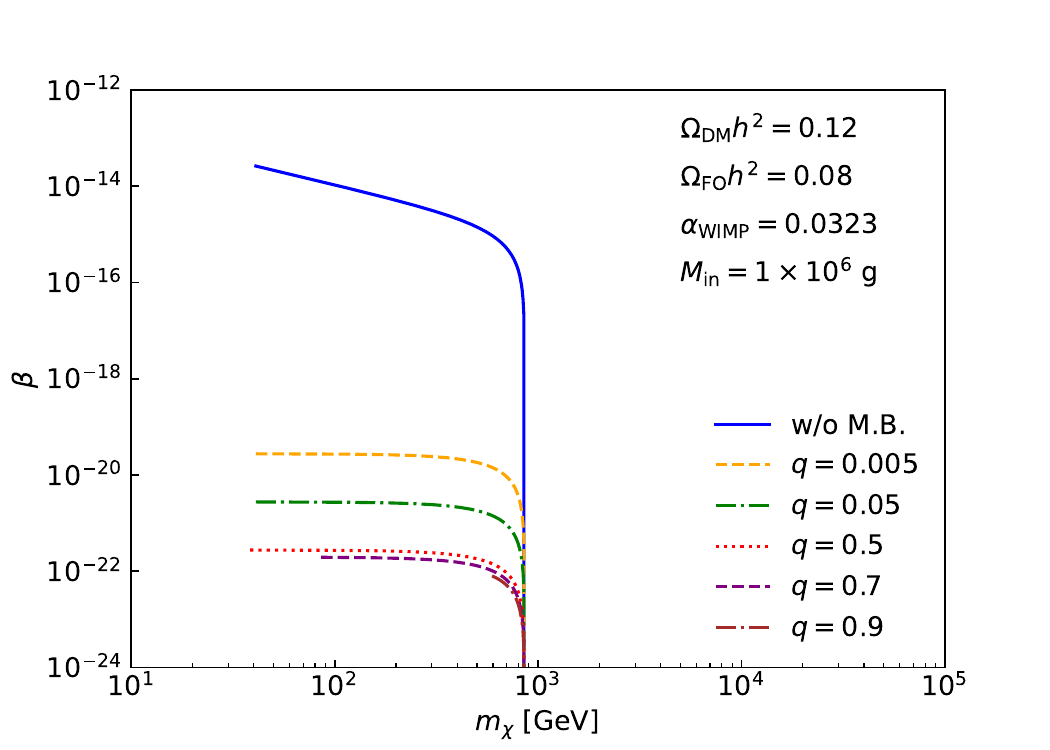}
\includegraphics[keepaspectratio, scale=0.35]{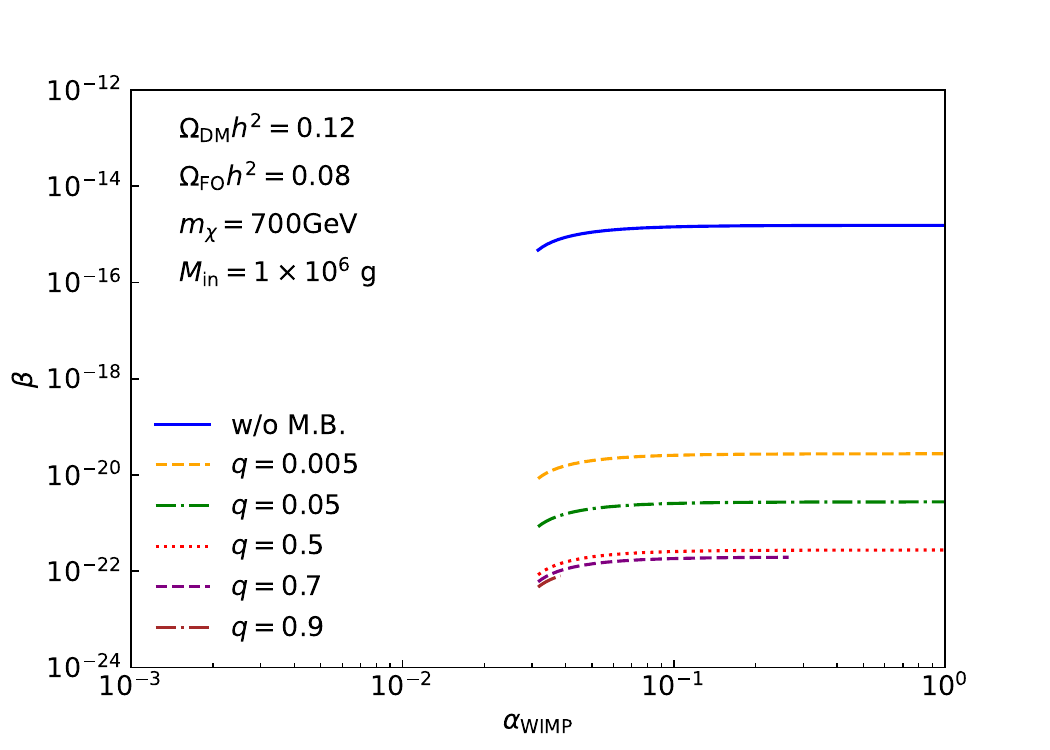}
 \caption{Constraints on initial PBH density, $\beta$, considering WIMPs and PBHs. Top: $\beta$ vs. $M_{\rm in}$ at various $q$. The left (right) gray region is excluded from the CMB (BBN) constraint. Bottom left: $\beta$  v.s, $m_\chi$. Bottom right: $\beta$ vs. $\alpha_{\rm WIMP}$.}
 \label{Fig:WIMP-PBH2}
 \end{figure}

Figure \ref{Fig:WIMP-PBH2} shows the constraint on the initial PBH density, $\beta$. The top panel shows $\beta$ against $M_{\rm in}$ at various $q$. The left (right) gray region is excluded from the CMB (BBN) constraint. The upper limit of $M_{\rm in}$ in each curve is due to the required $\beta < \beta_{\rm c} \propto 1/(q^{5/2+k}M_{\rm in}^{1+k})$. 

The bottom-left panel shows $\beta$ against $m_\chi$. As a strong memory burden effect yields long-lived PBHs, the initial PBH density, $\beta$, decreases with an increase in $q$. The upper limits of $m_\chi$ in the curves are obtained at $\Omega_{\rm DM}h^2 = \Omega_{\rm FO}h^2$. As we fix $\Omega_{\rm DM}h^2 = 0.12$ and $\alpha_{\rm WIMP}=0.0323$, the upper limit, $m_\chi = 858$ GeV, is obtained using Eq. (\ref{Eq:OmegaFOh2}). The lower limits of $m_\chi$ in the curves are controlled by the required $\beta < \beta_{\rm c}$. We obtain
\begin{equation}
\beta \simeq \frac{\sqrt{M_{\rm in}}}{4.2\times 10^{26} q\sqrt{M_{\rm P}}}\left( \Omega_{\rm DM}h^2 - 1.7\times 10^{-10} \frac{m_\chi^2}{\alpha_{\rm WIMP}^2} \right),
\label{Eq:betaWIMP}
\end{equation}
at $\Omega_{\rm FO}h^2 \gg \Omega_{\rm ev}h^2$ from Eq.(\ref{Eq:OmegaDMh2FO_and_PBH}). Thus, a small $m_\chi$ with a large $q$ increases $\beta$ and eliminates it via $\beta < \beta_{\rm c}$. 

The bottom-right panel shows $\beta$ according to the coupling of WIMPs, $\alpha_{\rm WIMP}$. The lower limit of $\alpha_{\rm WIMP}$ is caused by Eq. (\ref{Eq:OmegaFOh2}). The upper limits of $\alpha_{\rm WIMP}$ in the curves are obtained from the $\beta<\beta_{\rm c}$ constraint. Because $\beta \propto m_\chi^2/\alpha_{\rm WIMP}^2$,  the bottom-left ($\beta$ vs. $m_\chi$) and bottom-right ($\beta$ vs. $\alpha_{\rm WIMP}$)  panels are essentially reversed.

\begin{figure}[t]
\begin{center}
\includegraphics[keepaspectratio, scale=0.35]{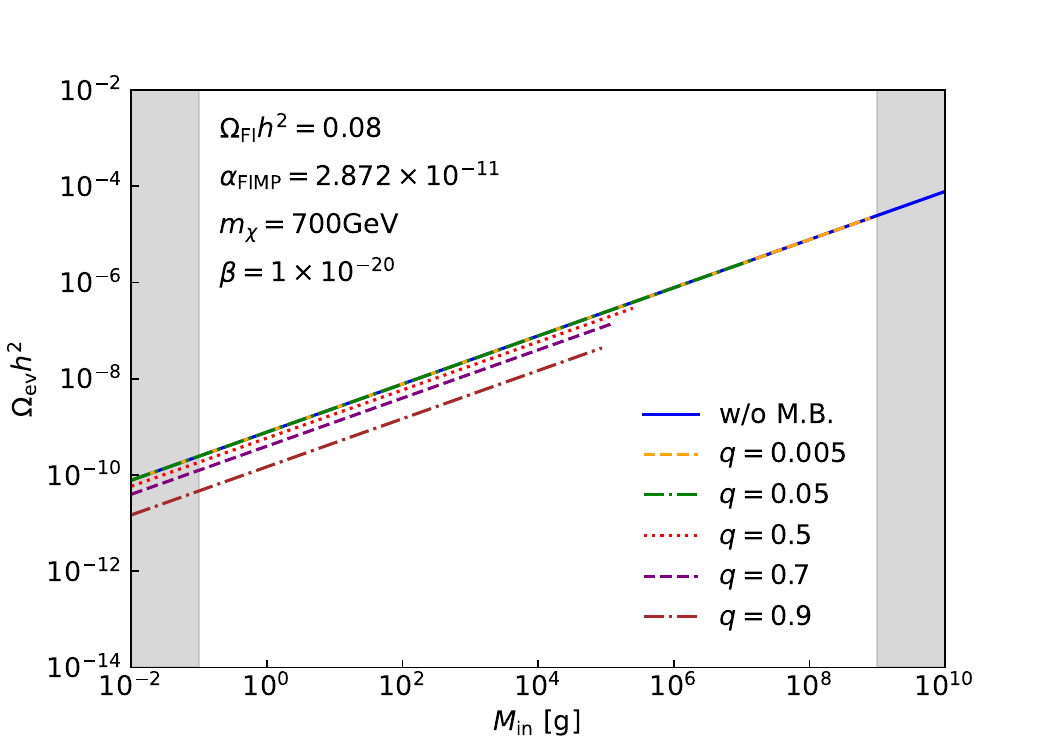}
\end{center}
\includegraphics[keepaspectratio, scale=0.35]{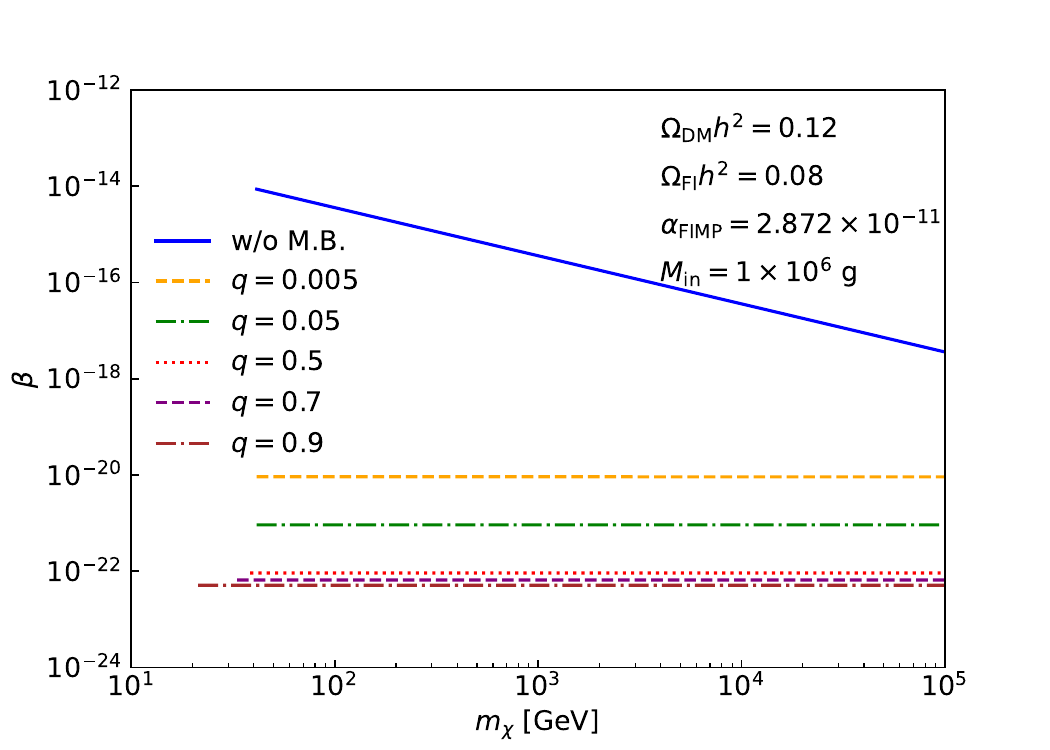}
\includegraphics[keepaspectratio, scale=0.35]{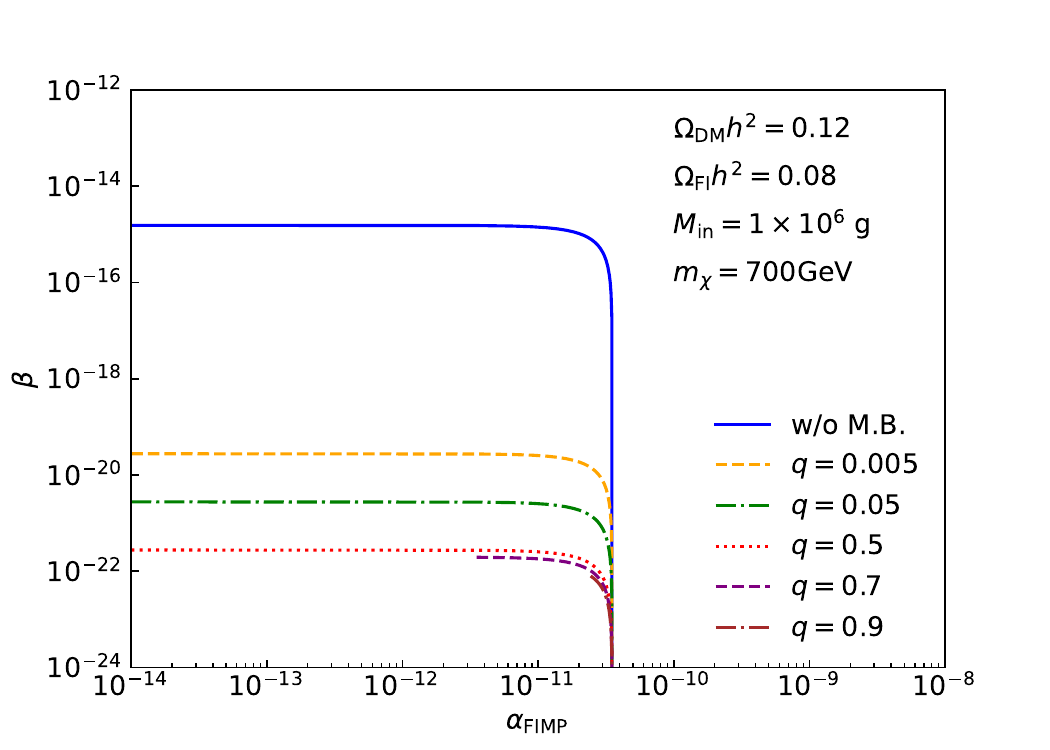}
 \caption{Relic abundance of DM and constraint on initial PBH density for FIMPs and PBHs. Only figures that differ from the case considering WIMPs are shown. Top: relic abundance of FIMPs owing to Hawking radiation from PBHs, $\Omega_{\rm ev}h^2$, according to initial PBH mass, $M_{\rm in}$. The left (right) gray region is excluded from the CMB (BBN) constraint. Bottom left: $\beta$ according to FIMP mass, $m_\chi$. Bottom right: $\beta$ according to FIMP coupling, $\alpha_{\rm FIMP}$.}
 \label{Fig:FIMP-PBH}
 \end{figure}

\subsection{FIMPs and PBHs}
Next, we consider DM consisting of FIMPs and PBHs. We assume that the masses of thermal bath particles are negligible compared with those of FIMPs and that the FI mechanism is described by pair production with a constant matrix element, $|\mathcal{M}(f_1f_2 \rightarrow \chi\chi)|^2 = \alpha_{\rm FIMP}^2$. For a simple model of FIMPs, the relic abundance of FIMPs, $\chi$, due to FI is \cite{DEramo2018JCAP,Gondolo2020PRD}
\begin{align}
\Omega_{\rm FI}h^2 =9.7 \times 10^{19} \alpha_{\rm FIMP}^2 .
\label{Eq:OmegaFIh2}
\end{align}
If the FI mechanism alone produces DM, then the observed relic abundance of DM, $\Omega_{\rm DM}h^2=0.12$, is consistent with $\alpha_{\rm FIMP} = 3.52\times 10^{-11}$ \cite{DEramo2018JCAP}. 

According to Ref. \cite{Cheek2022PRD}, in the case of the FI mechanism, particle interactions are small enough to avoid thermalization between particles from PBHs (FIMPs) and SM particles. If PBHs dominate the universe’s energy density before evaporating, then the annihilation cross section should be several orders of magnitude larger than that in the standard FI case to ensure efficient thermalization.

In this study, PBHs never dominate the energy density in the history of the universe. DM particles from PBH evaporation are expected not to interact with thermally produced FIMPs, as in the WIMP case. We consider the range of parameters $q$ at $\Omega_{\rm FI}h^2 \gg \Omega_{\rm ev}h^2$.

Given the feeble interactions of FIMPs, the FIMPs emitted by Hawking radiation can always contribute to the relic abundance of DM. The present-day DM density is 
\begin{align}
\Omega_{\rm DM}h^2 &= \Omega_{\rm FI}h^2 +  \Omega_{\rm PBH}h^2  + \Omega_{\rm ev}h^2 \nonumber \\
&= 9.7 \times 10^{19} \alpha_{\rm FIMP}^2 + 4.2 \times 10^{26} q \beta \left( \frac{M_{\rm P}}{M_{\rm in}} \right)^{1/2} \nonumber \\
& + 2.3 \times 10^5 C \beta \frac{m_\chi}{\rm GeV}, 
\label{Eq:OmegaDMh2FI_and_PBH}
\end{align}
where the first--third terms are due to FI, survived PBHs, and PBH evaporation.

As in the WIMP case, we consider $\Omega_{\rm FI}h^2 \gg \Omega_{\rm ev}h^2$. We regard $\alpha_{\rm FIIMP} = 2.872\times 10^{-11}$, $m_\chi = 700$ GeV ($\Omega_{\rm FI}h^2=0.08$), and $\beta = 1\times 10^{-20}$ as the benchmark. For this benchmark, we obtain the same prediction of $\Omega_{\rm DM}h^2$ with $M_{\rm in}$ (right panel of Fig. \ref{Fig:WIMP-PBH1}). Thus, we omit the plot of $\Omega_{\rm DM}h^2$ - $M_{\rm in}$ for the FIMP case. 

The top panel in Fig. \ref{Fig:FIMP-PBH} shows the relic abundance of FIMPs due to Hawking radiation from PBHs, $\Omega_{\rm ev}h^2$, according to the initial PBH mass, $M_{\rm in}$, at $\beta = 1\times 10^{-20}$ and $q=0.005-0.9$. $\Omega_{\rm FI}h^2 \gg \Omega_{\rm ev}h^2$ ($\Omega_{\rm ev}/\Omega_{\rm FI} \lesssim 10^{-4}$) holds. The left (right) gray region is excluded from the CMB (BBN) constraint.

The two bottom panels in Fig. \ref{Fig:FIMP-PBH} show the constraints on the initial PBH density, $\beta$, considering FIMPs. As in the WIMP case, a strong memory burden effect yields long-lived PBHs, and the initial PBH density, $\beta$, decreases with an increase in $q$ in the two bottom panels of Fig. \ref{Fig:FIMP-PBH}. 

The bottom-left panel shows $\beta$ according to the FIMP mass, $m_\chi$. Unlike in the WIMP case, the lower limits of $m_\chi$ in the curves are controlled by the WDM constraint (these lower limits are controlled by the $\beta < \beta_{\rm c}$ constraint in the WIMP case). From Eq.(\ref{Eq:OmegaDMh2FI_and_PBH}), we obtain
\begin{equation}
\beta \simeq \frac{\sqrt{M_{\rm in}}}{4.2\times 10^{26} q\sqrt{M_{\rm P}}}\left( \Omega_{\rm DM}h^2 - 9.7\times 10^{19} \alpha_{\rm FIMP}^2 \right),
\label{Eq:betaFIMP}
\end{equation}
at $\Omega_{\rm FI}h^2 \gg \Omega_{\rm ev}h^2$ in the FIMP case. Therefore, $\beta$ is independent of $m_\chi$ and the relation $\beta < \beta_{\rm c}$ does not constrain $m_\chi$. 

The bottom-right panel shows $\beta$ according to FIMP coupling, $\alpha_{\rm FIMP}$. Unlike in the WIMP case, the upper limit of $\alpha_{\rm FIMP}$ is caused by Eq. (\ref{Eq:OmegaFIh2}). The lower limit of  $\alpha_{\rm FIMP}$ is obtained through the $\beta < \beta_{\rm c}$ condition. A small $\alpha_{\rm FIMP}$ with a large $q$ increases $\beta$ and eliminates it via the $\beta < \beta_{\rm c}$ condition.

We would like to comment about gravitational production of DM from SM particles through graviton exchange. In addition to the thermal and PBH-induced production mechanisms discussed above, DM can in principle also be produced via FI by purely gravitational interactions through graviton exchange. This mechanism has been studied in detail in Ref.~\cite{Bernal2018PRD}. According to Ref.~\cite{Bernal2018PRD}, the thermally averaged cross section scales as ($n=2$ case) $\langle \sigma v \rangle \simeq T^2/M_{\rm P}^4$. The production rate is therefore $R(T) \sim n_{\rm SM}^2 \langle \sigma v \rangle \simeq T^8/M_{\rm P}^4$. During radiation domination, $H \sim T^2/M_{\rm P}$ and $s \sim T^3$, which gives $dY_\chi/dT \simeq T^2/M_{\rm P}^3$. Integrating up to the reheating temperature yields $Y_\chi \simeq T_{\rm RH}^3/M_{\rm P}^3$ and hence,
the resulting relic abundance scales as
\begin{equation}
\Omega_{\rm FI}^{\rm grav} h^2 \simeq \frac{m_\chi\, T_{\rm RH}^3}{M_{\rm P}^4}.
\end{equation}

In this study, we adopt a simple FIMP scenario, in which the FIMP relic abundance $\Omega_{\rm FI} h^2$ is given by Eq. (\ref{Eq:OmegaFIh2}). In the regime $\Omega_{\rm FI}^{\rm grav} h^2 \ll \Omega_{\rm FI} h^2$, namely 
\begin{equation}
m_\chi \ll \left( \frac{1.48 \times 10^{31} {\rm GeV}}{T_{\rm RH}}\right)^3 \alpha_{\rm FIMP}^2 ~ {\rm GeV},
\end{equation}
the contribution from graviton-mediated DM production is expected to be subdominant. Although the reheating temperature is model dependent, it is roughly estimated to lie in the range from a few GeV to $10^{15}$ GeV. Under these considerations, at least for $m_\chi \ll 3.2\times 10^{48} \alpha^2_{\rm FIMP}$ GeV, the graviton-mediated contribution is not expected to play a significant role in the present analysis. This condition is satisfied in the parameter region considered in this study.

\subsection{$k$-dependence}
So far, the analysis has not explicitly explored different values of the parameter $k$. Instead of varying $k$, we have focused on changes in the parameter $q$.  In the following, we discuss how varying $k$ may affect the parameter space and potentially modify the results.

The parameter $k$ enters the cosmological evolution primarily through the modification of the PBH evaporation timescale and temperature, $t_{\rm ev}(k)$ and $T_{\rm ev}(k)$, as discussed in Sec.~2. At the level of parametric dependence, increasing $k$ delays the evaporation and lowers $T_{\rm ev}$, thereby increasing the typical momentum of DM particles emitted at evaporation relative to the plasma temperature.

This effect propagates into the relic abundance and the WDM constraint mainly through $T_{\rm ev}$. For instance, the free-streaming length and the corresponding lower bound on the DM mass scale approximately as $m_{\chi}^{\rm WDM} \propto T_{\rm ev}^{-1}(k)$, implying that a larger value of $k$ strengthens the WDM constraint.  Apart from this indirect dependence, no additional qualitative modification of the production mechanism arises from varying $k$ within the range of interest. Quantitatively, increasing $k$ shifts the WDM excluded region toward larger DM masses through the reduction of $T_{\rm ev}$, while leaving the structure of the allowed parameter space unchanged.

In the present work, we therefore fix $k$ and instead vary the parameter $q$, which directly controls the strength of the memory burden effect, in order to efficiently illustrate its phenomenological impact. A systematic scan over $k$ would lead to a quantitative reshaping of the parameter space but would not alter the qualitative conclusions of our analysis, and we leave such a study for future work.

\section{Summary \label{section:summary}}
The lifetime of PBHs is extended by the memory burden effect. Hence, light PBHs may survive and become candidates for DM. We assume that DM is made of thermally produced WIMPs/FIMPs, WIMPs/FIMPs produced via the Hawking radiation of PBHs, and PBHs that survived via the memory burden effect. We focus on a case where the effect of the thermal production of WIMPs/FIMPs substantially exceeds that of PBH particle production via Hawking radiation.

Focusing on parameter regions in which PBHs never dominate the energy density of the Universe, we derived the relic abundance of DM and identified a sufficient condition under which DM particles emitted by PBHs do not enter thermal equilibrium with the thermal bath. In this regime, the contribution from PBH evaporation remains independent of the standard FO (or FI) process, and the total relic abundance of DM can be consistently obtained as a sum of the three components.

In addition, we have examined the potential contribution from gravitational FI via graviton exchange and demonstrated that it remains subdominant within the parameter space considered in this work.  

A more detailed numerical investigation of scenarios close to thermalization boundaries or a systematic scan over the full memory burden parameter space would be an interesting direction for future work.

\section*{Acknowledgement}
The author, T. K.,  thanks Anish Ghoshal for useful discussions.

\vspace{3mm}




\begin{thebibliography}{0}    

\bibitem{Arbey2021PPNP}
A. Arbey and F. Mahmoudi, \Journal{\PPNP}{119}{103865}{2021}.

\bibitem {Kolb1990text}
E. W. Kolb and M. S. Turner, {\it The Early Universe}, Frontiers in Physics. 69, 1 (1990).

\bibitem{Hall2010JHEP}
L. J. Hall, K. Jedamzik, J. March-Russell, and S. M. West, \Journal{\JHEP}{03}{080}{2010}.

\bibitem{Bernal2017IJMPA}
N. Bernal, M. Heikinheimo, T. Tenkanen, K. Tuominen, and Vaskonen, \Journal{\IJMPA}{32}{2017}{1730023}.

\bibitem{Carr1975APJ}
B. J. Carr, \Journal{\APJ}{201}{1}{1975}.

\bibitem{Carr2020ARNPS}
B. Carr and F. K{\" u}hnel, \Journal{\ARNPS}{70}{355}{2020}.

\bibitem{Carr2021RPP}
B. Carr, K. Kohri, Y. Sendouda, and J. Yokoyama, \Journal{\RPP}{84}{116902}{2021}.

\bibitem{Auffinger2023PPNP}
J. Auffinger, \Journal{\PPNP}{131}{104040}{2023}.


\bibitem{Khlopov2010RAA}
M. Yu. Khlopov. Res. Astron. Astrophys. (2010) V.10,PP. 495-528.

\bibitem{Belotsky2014MPLA}
K. M. Belotsky, A. D. Dmitriev, E. A. Esipova, V. A. Gani, A. V. Grobov, M. Yu. Khlopov, A. A. Kirillov, S. G. Rubin, and I. V. Svadkovsky, \Journal{\MPLA}{29}{1440005}{2014}.
 
\bibitem{Belotsky2019EPJC}
K. M. Belotsky, V. I. Dokuchaev, Yu. N. Eroshenko, E. A. Esipova, M. Yu. Khlopov, L. A. Khromykh, A. A. Kirillov, V. V. Nikulin, S. G. Rubin, and I. V. Svadkovsky, \Journal{\EPJC}{79}{246}{2019}.


\bibitem{Heydari2022EPJC}
S. Heydari and K. Karami, \Journal{\EPJC}{82}{83}{2022}.

\bibitem{Heydari2022JCAP}
S. Heydari and K. Karami, \Journal{\JCAP}{03}{033}{2022}.

\bibitem{Heydari2024JCAP}
S. Heydari and K. Karami, \Journal{\JCAP}{02}{047}{2024}.

\bibitem{Heydari2024EPJC}
S. Heydari and K. Karami, \Journal{\EPJC}{84}{127}{2024}.

\bibitem{Heydari2024ApJ}
S. Heydari and K. Karami, ApJ 975 148 (2024), \Journal{\APJ}{975}{148}{2024}

\bibitem{Cheek2022PRD1}
A. Cheek, L, Heurtier, Y. F. Perez-Gonzalez, and J. Turner, \Journal{\PRD}{105}{015022}{2022}.

\bibitem{Hawking1975CMP}
S. W. Hawking, \Journal{\CMP}{43}{199}{1975}.

\bibitem{Dvali2016FortschrPhys}
G. Dvali, Fortschr. Phys. 64, 106 (2016).

\bibitem{Dvali2020PRD}
G. Dvali, L. Eisemann, M. Michel, and S. Zell, \Journal{\PRD}{102}{103523}{2020}.

\bibitem{Alexandre2024PRD}
A. Alexandre, G. Dvali, and E. Koutsangelas, \Journal{\PRD}{110}{036004}{2024}.

\bibitem{Dvali2024PRD}
G. Dvali, J. S. Valbuena-Berm{\' u}dez, and M. Zantedeschi, \Journal{\PRD}{110}{056029}{2024}.

\bibitem{Thoss2024MNRAS}
V. Thoss, A. Burkert, and K. Kohri, \Journal{\MNRAS}{532}{451}{2024}.

\bibitem{Chianese2025arXiv}
M. Chianese, arXiv:2504.03838 (2025).

\bibitem{Haque2024JCAP}
M. R. Haque, S. Maity, D. Maity, and Y. Mambrini, \Journal{\JCAP}{07}{002}{2024}.

\bibitem{Montefalcone2025arXiv}
G. Montefalcone, D. Hooper, K. Freese, C. Kelso, F. K{\" u}hnel, and P. Sandick, arXiv:2503.21005 (2025).

\bibitem{Dvali2025arXiv}
G. Dvali, M. Zantedeschi, and S. Zell, arXiv:2503.21740 (2025).

\bibitem{KohriPRD2025}
K. Kohri, T. Terada, and T. T. Yanagida,  \Journal{\PRD}{111}{063543}{2025}.

\bibitem{Bhaumik2024JHEP}
N. Bhaumik, M. R. Haque, R. K. Jain, and M. Lewicki, \Journal{\JHEP}{10}{142}{2024}.

\bibitem{Barman2024JCAP09}
B. Barman, M. R. Haque, and \'{O}. Zapata, \Journal{\JCAP}{09}{020}{2024}.

\bibitem{Barman2024JCAP10}
B. Barman, K. Loho, and \'{O}. Zapata, \Journal{\JCAP}{10}{065}{2024}.

\bibitem{Balaji2024JCAP11}
S. Balaji, G. Dom\`{e}nech, G. Franciolini, A. Ganz, and J. Tr\"{a}nkle, \Journal{\JCAP}{11}{026}{2024}.

\bibitem{Jiang2024JCAP12}
Y. Jiang, C. Yuan, C.-Z. Li, and Q.-G. Huang, \Journal{\JCAP}{12}{016}{2024}.

\bibitem{Loc2025PRD}
N. P. D. Loc, \Journal{\PRD}{111}{023509}{2025}.

\bibitem{Athron2025JCAP}
P. Athron, M. Chianese, S. Datta, R. Samanta, and N. Saviano, \Journal{\JCAP}{05}{005}{2025}.

\bibitem{Barker2025PRD}
W. Barker, B. Gladwyn, and S. Zell, \Journal{\PRD}{111}{123033}{2025}.

\bibitem{Gross2025arXiv}
M. Gross, Y. Mambrini, and M. R. Haque, arXiv:2509.02701.

\bibitem{Bandyopadhyay2025arXiv}
D. Bandyopadhyay, D. Borah, and N. Das, arXiv:2501.04076 (2025).

\bibitem{Barman2024arXiv}
B. Barman, K. Loho, and O. Zapata, arXiv:2412.13254 (2024).

\bibitem{Calabrese2025arXiv}
R. Calabrese, M. Chianese, and N. Saviano, \Journal{\PRD}{111}{083008}{2025}.

\bibitem{Borah2024arXiv}
D. Borah and N. Das, arXiv:2410.16403 (2024).

\bibitem{Chianese2024arXiv}
M. Chianese, A. Boccia, F. Iocco, G. Miele, and N. Saviano, arXiv:2410.07604 (2024).

\bibitem{Chaudhuri2025arXiv}
A. Chaudhuri, K. Pal, and R. Mohanta, arXiv:2025.09153 (2025).

\bibitem{Dondarini2025arXiv}
A. Dondarini, G. Marino, P. Panci, and M. Zantedeschi, arXiv:2506.13861(2025).

\bibitem{Zantedeschi2025arXiv}
M. Zantedeschi and L. Visinelli, arXiv:2410.07037 (2025)

\bibitem{Basumatary2025PRD}
U. Basumatary, N. Raj, and A. Ray, \Journal{\PRD}{111}{L041306}{2025}.

\bibitem{Federico2025PRD}
K. Federico and S. Profumo, \Journal{\PRD}{111}{063006}{2025}.

\bibitem{Gondolo2020PRD}
P. Gondolo, P. Sandick, and B. S. E. Haghi, \Journal{\PRD}{102}{0595018}{2020}.

\bibitem{Fujita2014PRD}
T. Fujita, K. Harigaya, M. Kawasaki, and R. Matsuda, \Journal{\PRD}{89}{103501}{2014}.

\bibitem{Cheek2022PRD}
A. Cheek, L, Heurtier, Y. F. Perez-Gonzalez, and J. Turner, \Journal{\PRD}{105}{015023}{2022}.

\bibitem{Chanda2025arXiv}
P. Chanda, S. Mukherjee, and J. Unwin, arXiv:2505.02935 (2025).

\bibitem{Kitabayashi2021IJMPA}
T. Kitabayashi, \Journal{\IJMPA}{36}{2150139}{2021}.

\bibitem{Kitabayashi2022PTEP}
T. Kitabayashi, \Journal{\PTEP}{2022}{033B02}{2022}.

\bibitem{Kitabayashi2022PTEP2}
T. Kitabayashi, \Journal{\PTEP}{2022}{123B01}{2022}.

\bibitem{Kitabayashi2022IJMPA}
T. Kitabayashi, \Journal{\IJMPA}{37}{2250181}{2022}.

\bibitem{Kitabayashi2024PDU}
T. Kitabayashi, \Journal{\PDU}{45}{101506}{2024}.

\bibitem{Takeshita2025IJMPA}
A. Takeshita and T. Kitabayashi, \Journal{\IJMPA}{40}{2550036}{2025}.

\bibitem{Mambrini2021text}
Y. Mambrini, {\it Particles in the Dark Universe} , Springer Ed. (2021).

\bibitem{Kohri1999PRD}
K. Kohri and J. Yokoyama, \Journal{\PRD}{61}{023501}{1999}.

\bibitem{Kawasaki2000PRD}
M. Kawasaki, K. Kohri, and N. Sugiyama, \Journal{\PRD}{62}{023506}{2000}.

\bibitem{Carr2010PRD}
B. Carr, K. Kohri, Y. Sendouda, and J. Yokoyama, \Journal{\PRD}{81}{104019}{2010}.

\bibitem{Tomita1967PTP}
K. Tomita, \Journal{\PTP}{37}{831}{1967}.

\bibitem{Matarrese1993PRD}
S. Matarrese, O. Pantano, and D. Saez, \Journal{\PRD}{47}{1311}{1993}.

\bibitem{Matarrese1994PRL}
S. Matarrese, O. Pantano, and D. Saez, \Journal{\PRL}{72}{320}{1994}.

\bibitem{Matarrese1998PRD}
S. Matarrese, S. Mollerach, and M. Bruni, \Journal{\PRD}{58}{043504}{1998}.

\bibitem{Noh2004PRD}
H. Noh and J. -c. Hwang, \Journal{\PRD}{69}{104011}{2004}.

\bibitem{Carbone2005PRD}
C. Carbone and S. Matarrese, \Journal{\PRD}{71}{043508}{2005}.

\bibitem{Nakamura2007PTP}
K. Nakamura,  \Journal{\PTP}{117}{17}{2007}.

\bibitem{Ananda2007PRD}
K. N. Ananda, C. Clarkson, and D. Wands, \Journal{\PRD}{75}{123518}{2007}.

\bibitem{Baumann2007PRD}
D. Baumann, P. J. Steinhardt, K. Takahashi, and K. Ichiki, \Journal{\PRD}{76}{084019}{2007}.

\bibitem{Osano2007JCAP}
B. Osano, C. Pitrou, P. Dunsby, J.-P. Uzan, and C. Clarkson, \Journal{\JCAP}{04}{003}{2007}.

\bibitem{Saito2009PRL}
R. Saito and J. Yokoyama, \Journal{\PRL}{102}{161101}{2009}. [Erratum ibid. {\bf 107}, 069901(2011)].

\bibitem{Espinosa2018JCAP}
J. R. Espinosa, D. Racco, and A. Riotto, \Journal{\JCAP}{09}{012}{2018}.

\bibitem{Kohr2018PRD}
K. Kohri and T. Terada, \Journal{\PRD}{97}{123532}{2018}.

\bibitem{Domenech2020IJMPD}
G. Dom{\` e}nech, \Journal{\IJMPD}{29}{2050028}{2020}.

\bibitem{Domenech2020JCAP}
G. Dom{\` e}nech, S. Pi, and M. Sasaki, \Journal{\JCAP}{08}{017}{2020}.

\bibitem{Wang2023arXiv}
X. Wang, Y.-l. Zhang, R. Kimura, and M. Yamaguchi, arXiv:2209.12911 (2023).

\bibitem{Choudhury2014PLB}
S. Choudhury and A. Mazumdar, \Journal{\PLB}{733}{270}{2014}.

\bibitem{Choudhury2023arXiv}
S. Choudhury, A. Karde, S. Panda, and M. Sami, arXiv:2306.12334 (2023).

\bibitem{Domenech2021Universe}
G. Dom{\` e}nech, \Journal{\UNIV}{7}{398}{2021}.

\bibitem{Inomata2019JCAP}
K. Inomata, K. Kohri, T. Nakama and T. Terada, \Journal{\JCAP}{10}{071}{2019}.

\bibitem{Inomata2020PRD}
K. Inomata, M. Kawasaki, K. Mukaida, T. Terada, and T. T. Yanagida, \Journal{\PRD}{101}{123533}{2020}.

\bibitem{Papanikolaou2021JCAP}
T. Papanikolaou, V. Vennin, and D. Langlois, \Journal{\JCAP}{03}{053}{2021}.

\bibitem{Bhaumik2022JHEP}
N. Bhaumik, A. Ghoshal, and Marek Lewicki, \Journal{\JHEP}{07}{130}{2022}.

\bibitem{Inomata2019PRD}
K. Inomata, K. Kohri, T. Nakama, and T. Terada, \Journal{\PRD}{100}{043532}{2019}.

\bibitem{Domenech2021JCAP}
G. Dom{\` e}nech, C. Lin, and M. Sasaki, \Journal{\JCAP}{04}{062}{2021}. [Erratum ibid. {\bf 11} (2021) E01].

\bibitem{Domenech2021PLB}
G. Dom{\` e}nech, V. Takhistov, and M. Sasaki, \Journal{\PLB}{823}{136722}{2021}.

\bibitem{Borah2023JHEP}
D. Borah, S. J. Das, R. Samanta, and F. R. Urban, \Journal{\JHEP}{03}{127}{2023}.

\bibitem{KawasakiArXiv2023}
M. Kawasaki and K. Murai, arXiv:2308:13134 (2023).

\bibitem{LIGO2010CQG}
LIGO Scientific Collaboration, \Journal{\CQG}{27}{084006}{2010}.

\bibitem{LIGO2015CQG}
LIGO Scientific Collaboration, \Journal{\CQG}{32}{074001}{2015}.

\bibitem{VIRGO2015CQG}
VIRGO Collaboration, \Journal{\CQG}{32}{024001}{2015}.

\bibitem{KAGRA2012CQG}
KAGRA Collaboration, \Journal{\CQG}{29}{124007}{2012}.

\bibitem{Punturo2010CQG}
M. Punturo, et al., \Journal{\CQG}{27}{194002}{2010}.

\bibitem{Maggiore2020JCAP}
M. Maggiore, et al., \Journal{\JCAP}{03}{050}{2020}.

\bibitem{LIGO2017CQG}
LIGO Scientific Collaboration, \Journal{\CQG}{34}{044001}{2017}.

\bibitem{LISA2017arXiv}
LISA Collaboration,  arXiv:1702.00786 (2017).

\bibitem{Seto2001PRL}
N. Seto, S. Kawamura, and T. Nakamura,  \Journal{\PRL}{87}{221103}{2001}.

\bibitem{Kawamura2006CQG}
S. Kawamura, et al., \Journal{\CQG}{23}{S125}{2006}.

\bibitem{Crowder2005PRD}
J. Crowder and N. J. Cornish, \Journal{\PRD}{72}{083005}{2005}.

\bibitem{Harry2006CQG}
G. M. Harry, P. Fritschel, D. A. Shaddock, W. Folkner, and E. S. Phinney, \Journal{\CQG}{23}{4887}{2006} [Erratum ibid. 23 (2006) 7361].

\bibitem{Hui1997ApJ}
L. Hui, N. Y. Gnedin, and Y. Zhang, \Journal{\APJ}{486}{599}{1997}.

\bibitem{Gnedin2002MNRAS}
N. Y. Gnedin and A. J. S. Hamilton, \Journal{\MNRAS}{334}{107}{2002}.

\bibitem{Baldes2020JCAP}
I. Baldes, Q. Decant, D. C. Hooper, and L. Lopez-Honorez, \Journal{\JCAP}{08}{045}{2020}.

\bibitem{Boyarsky2009JCAP}
A. Boyarsky, J. Lestourgues, O. Ruchayskiy, and M. Viel, \Journal{\JCAP}{05}{012}{2009}.

\bibitem{Baur2017JCAP}
J. Baur, N. P.-Delabrouille, C. Y{\` e}che, A. Boyarsky, O. Ruhayskiy, {\' E}. Armengaud, and J. Lesgourgues, \Journal{\JCAP}{12}{013}{2017}.

\bibitem{PDG}
S. Navas et al. (Particle Data Group), \Journal{\PRD}{110}{030001}{2024}.

\bibitem{Irsic2017PRD}
V. Ir{\v s}i{\v c}, et al., \Journal{\PRD}{96}{032522}{2017}.

\bibitem{Masina2020EPJP}
I. Masina, \Journal{\EPJP}{135}{552}{2020}.

\bibitem{Planck2018AA}
N. Aghanim, et al., (Planck Collaboration), \Journal{\AandA}{641}{A6}{2020}.

\bibitem{DEramo2018JCAP}
F. D'Eramo, N. Fernandez, and S. Profumo, \Journal{\JCAP}{02}{046}{2018}.

\bibitem{Fujii2002PRD}
M. Fujii, K. Hamaguchi, and T. Yanagida, \Journal{\PRD}{65}{115012}{2002}.

\bibitem{Lennon2018JCAP}
O. Lennon, J. March-Russell, R. Petrossian-Byrne, and H. Tillim, \Journal{\JCAP}{04}{009}{2018}.

\bibitem{Bernal2018PRD}
N. Bernal, M. Dutra, Y. Mambrini, K. Olive, M. Peloso, and M. Pierre, \Journal{\PRD}{97}{11520}{2018}.

\end{thebibliography}




\end{document}